\documentclass[twocolumn,preprintnumbers,amsmath,amssymb,floatfix,superscriptaddress,nofootinbib]{revtex4}

\usepackage[paperwidth=210mm,paperheight=297mm,centering,hmargin=1.7cm,vmargin=2.2cm]{geometry}

\usepackage{amssymb,amsmath,latexsym,
multirow,comment,appendix,slashed,color,afterpage,multirow,makecell}

\usepackage[colorlinks=true
,urlcolor=blue
,anchorcolor=blue
,citecolor=blue
,filecolor=blue
,linkcolor=blue
,menucolor=blue
,linktocpage=true
,pdfproducer=medialab
,pdfa=true
]{hyperref}

\pdfoutput=1
\allowdisplaybreaks
\usepackage{bm}
\usepackage{setspace}
\usepackage{tikz}
\usetikzlibrary{calc}

\usepackage{cleveref}
\crefname{table}{Table}{Tables}
\crefname{equation}{Eq.}{Eqs.}
\crefname{appendix}{App.}{Apps.}
\crefname{section}{Sec.}{Secs.}
\crefname{figure}{Fig.}{Figs.}
\usepackage[normalem]{ulem}

\def\eg{\textit{e.g.}}

\usepackage{dcolumn}
\usepackage{bm}
\usepackage{color}
\usepackage{xspace}

\newcommand{\pt}{$p_t$~}

\newcommand*{\PYTHIA}{\textsc{Pythia}\xspace}
\newcommand*{\Pythia}{\PYTHIA}
\newcommand*{\kt}{\ensuremath{k_{t}}}
\newcommand*{\lambdaD}{\ensuremath{\Lambda_\text{D}}}
\newcommand{\s}{\hspace{0.8pt}}

\begin{document}
\preprint{CERN-TH-2023-007}

\vspace*{10pt}

\title{Dark Sector Showers in the Lund Jet Plane}

\author{Timothy Cohen}
\affiliation{Theoretical Physics Department, CERN, 1211 Geneva, Switzerland \vspace{3pt}}
\affiliation{Theoretical Particle Physics Laboratory, EPFL, 1015 Lausanne, Switzerland \vspace{3pt}}
\affiliation{Institute for Fundamental Science, University of Oregon, Eugene, OR 97403, USA \vspace{3pt}}

\author{Jennifer Roloff}
\affiliation{Physics Department, Brookhaven National Laboratory, Upton, NY 11973-5000, USA\vspace{3pt}}

\author{Christiane Scherb}
\affiliation{Physics Division, Lawrence Berkeley National Laboratory, Berkeley, CA 94720, USA \vspace{3pt}}
\affiliation{Berkeley Center for Theoretical Physics, University of California, Berkeley, CA 94720, USA \vspace{3pt}}

\begin{abstract}
\begin{center}
{\bf Abstract}
\vspace{-7pt}
\end{center}
We investigate the consequences of models where dark sector quarks could be produced at the LHC, which subsequently undergo a dark parton shower, generating jets of dark hadrons that ultimately decay back to Standard Model hadrons.  This yields collider objects that can be nearly indistinguishable from Standard Model jets, motivating the reliance on substructure observables to tease out the signal.  However, substructure predictions are sensitive to the details of the incalculable dark hadronization.  We show that the Lund jet plane provides a very effective tool for designing observables that are resilient against the unknown impact of dark hadronization on the substructure properties of dark sector jets.  
\end{abstract}

\maketitle

\begin{spacing}{1.1}
\section{Introduction}
\noindent
Although there is overwhelming evidence for the existence of dark matter, we do not yet know the detailed properties of this beyond-the-Standard-Model state(s).  One minimal assumption is that it is an additional weakly interacting state.  However, given the complexities of the Standard Model itself, one should be open to the possibility that the dark matter is one or more stable particles that emerge from a dark sector with non-trivial dynamics.  In this letter, we consider the case where the dark sector involves a new confining force that binds a set of dark quarks into dark hadrons, in close analogy with quantum chromodynamics (QCD).  In a wide class of models, some of these dark hadrons could be stable, and could therefore provide viable dark matter candidates.  We also expect the relic density of the dark matter to be generated by some non-trivial couplings to the Standard Model bath in the early universe.  This motivates studying models where the dark sector is connected to the Standard Model by a so-called portal interaction.  This paradigm has received tremendous attention in recent years; our interest here is exploring the implications for Large Hadron Collider (LHC) phenomenology, \eg~see the Snowmass study~\cite{Albouy:2022cin}.

Models with QCD-like strong dynamics in the dark sector connected to the Standard Model by a portal can result in a wide range of signatures at the LHC.  Following the pioneering papers on Hidden Valleys~\cite{Strassler:2006im, Strassler:2006ri, Han:2007ae}, a number of other classes of signatures have been identified, \eg~lepton jets~\cite{Arkani-Hamed:2008kxc, Baumgart:2009tn, Chan:2011aa, ATLAS:2015itk, Buschmann:2015awa, ATLAS:2019tkk, duPlessis:2021xuc}, emerging jets~\cite{Schwaller:2015gea, CMS:2018bvr, Renner:2018fhh}, semi-visible jets~\cite{Cohen:2015toa, Cohen:2017pzm, Beauchesne:2017yhh, Kar:2022hxn, Cazzaniga:2022hxl, Beauchesne:2022phk}, and soft bombs~\cite{Harnik:2008ax, Knapen:2016hky}.
The properties of the dark sector can be parameterized by combination of physically meaningful choices: a confinement scale $\Lambda_\text{D}$, the number of dark colors $n_C$, and the number of dark quark flavors $n_F$.  In order to make predictions for the LHC, one must also model dark hadronization, 
which introduces a number of additional model parameters.
Finally, one must specify the portal, which could utilize a new coupling to a Standard Model operator~\cite{Holdom:1985ag,Patt:2006fw,Falkowski:2009yz}, or rely on the introduction of a new physics mediator \cite{Knapen:2021eip,Ismail:2017ulg,Frandsen:2012rk,Buchmueller:2013dya,Dreiner:2013vla,Buchmueller:2014yoa,Hamaguchi:2014pja,Harris:2014hga,Jacques:2015zha,Liew:2016oon,Englert:2016joy,Bernreuther:2019pfb,Cheng:2019yai,Cohen:2017pzm,Agrawal:2014aoa,Jubb:2017rhm,Blanke:2017fum,Renner:2018fhh}. For simplicity, we will model the production of dark quarks using a contact operator in the studies presented below~\cite{Goodman:2010yf,Beltran:2010ww,Fox:2011pm}.

ATLAS and CMS have recently performed searches for models that involve dark showers~\cite{CMS:2021dzg,ATLAS:2022aaw,CMS:2018bvr}, and there have been a number of proposals for how to extend the reach of these searches using jet substructure~\cite{Park:2017rfb, Kar:2020bws, Canelli:2021aps, Faucett:2022zie}. However, many of these observables are highly sensitive to poorly-understood effects, particularly the dark sector hadronization, see~\cite{Cohen:2020afv} for a study that characterized the size of theory errors for one class of substructure observables.  These effects are not easily parameterized, and there are large uncertainties associated to their modeling. Searches designed for a single choice of hadronization modeling may lose a lot of sensitivity with other plausible hadronization options. Even worse than losing sensitivity, the limits may not even apply to the true prediction of the theory, depending on how the signal has been parameterized.  Approaches to mitigate this issue are clearly of great value.

There are many jet substructure observables that are well known to be sensitive to the non-perturbative dynamics of confinement.  For example, one can simply count the number of constituents in a jet: this observable depends significantly on the model for fragmentation and hadronization.  Such modeling can be tuned against data for the Standard Model, but this is not (yet) possible for a dark sector.  
In contrast, we will demonstrate that the Lund jet plane (LJP) provides a very useful tool to accomplish the goal of minimizing the sensitivity to the underlying hadronization model, while simultaneously providing a powerful discriminator.  The LJP provides an intuitive way of factorizing different physical effects in QCD~\cite{Dreyer:2018nbf}, extending the concept of the Lund diagrams~\cite{Andersson:1988gp} to an experimentally viable observable.  As described in detail below, the LJP is constructed to separate the parton shower contribution to the jet from the hadronization effects.  Measurements of the LJP have demonstrated the experimental power of the LJP to explore perturbative and non-perturbative QCD effects \cite{ATLAS:2020bbn,ALICE:2021yet}, and it has also been proposed as a tagger for identifying different types of jets~\cite{Khosa:2021cyk,Khosa:2021zbx,Fedkevych:2022mid,Dreyer:2020brq}. In this paper, we explore how these same insights can be used to design interpretable observables to tag dark sector jets, as well as to study the modeling of dark showers across different simulation platforms.  

The rest of this letter is organized as follows.  First, we describe the details of the simulation and the algorithm for calculating the LJP in \cref{setup}.  We then apply this to a variety of dark sector scenarios in \cref{results}.  We explore the resilience against variation in the hadronization parameters in \cref{robust}.  Finally, we provide conclusions and future directions in \cref{conclusions}.

\section{Tools}
\label{setup}
\noindent 
In this section, we explain the details of the simulations including our benchmark parameter choices, and the computation of the LJP.
The dark sector simulation is performed using the Hidden Valley \Pythia module~\cite{Sjostrand:2014zea,Carloni:2010tw}.  Our benchmark parameter point is as follows: number of colors $n_C = 3$, number of dark quark flavors $n_F = 3$, dark confinement scale $\lambdaD = 5 \text{ GeV}$, dark quark mass $m_{Q_\text{D}} = 2.5 \text{ GeV}$, dark pion mass $m_{\pi_\text{D}} = 5 \text{ GeV}$, dark rho mass $m_{\rho_\text{D}} = 5 \text{ GeV}$.  For the hadronization and fragmentation parameters, we take the default choices (see the Hidden Valley section of the \Pythia manual) $a_L = 0.3$, $b_{m^2_{Q_\text{D}}} = 0.8$, $r_{Q_\text{D}} = 1$, and fraction of $\rho_\text{D} = 0.75$.  Other benchmark points are provided in \cref{tab:Rinv}.

Proton-proton collisions are simulated at $\sqrt{s} = 13 \text{ TeV}$.  We pair produce dark quarks through a contact operator portal connecting them to the Standard Model quarks.  We stop the simulation at five different stages: (1) we produce the dark sector partons and simulate their subsequent parton shower, so the final state at this stage are dark sector partons; (2) we hadronize the dark sector partons into dark hadrons; (3) we decay the dark hadrons into Standard Model partons; (4) we shower the Standard Model partons that were produced in the decay, generating a final state with many Standard Model partons; (5) we hadronize the Standard Model final state.  
Isolating these different stages allows us to explore the impact of each effect on the structure of the LJP.\footnote{Note that we assume all of the dark hadrons are unstable. Including some fraction of stable dark matter candidates thereby producing a ``semi-visible jet'' would not change any of our conclusions as long as a substantial number of Standard Model hadrons appear in the final state.}

At each stage of the simulation, large-radius ($R=1.0$) anti-$k_t$ jets~\cite{Cacciari:2008gp} are clustered from the partons or hadrons using Fastjet~\cite{Cacciari:2011ma,Cacciari:2005hq}. Jets are required to have \pt $> 1000$~GeV.  
Given the set of jets that are simulated at a given stage, we then compute the associated LJP.
To reconstruct the primary LJP, each reconstructed jet is reclustered using the Cambridge/Aachen (C/A) algorithm \cite{Dokshitzer:1997in,Wobisch:1998wt}. The final clustering step of this C/A jet is undone, resulting in two pseudojets, and the lower-\pt pseudojet is assumed to be an emission off of the full jet. Then, the higher-\pt pseudojet is considered to be the core of the remaining jet, and the declustering proceeds with this pseudojet. In this case, we compute the transverse momentum of emission with respect to the core ($k_{t}$), and the angular distance between the emission and the core ($\Delta R = \sqrt{\Delta \eta^2 + \Delta \phi^2}$). This process continues until the declustering reaches the end of the clustering procedure. The results of this declustering can then be plotted, with each emission from the core contributing one point on the LJP. In general, each jet will have several emissions, and when normalized to the number of jets, the integral of the LJP corresponds to the average number of total emissions across all jets. 

\section{Results}
\label{results}
\begin{figure}[thb]
  \centering
  \includegraphics[width=0.32\textwidth]{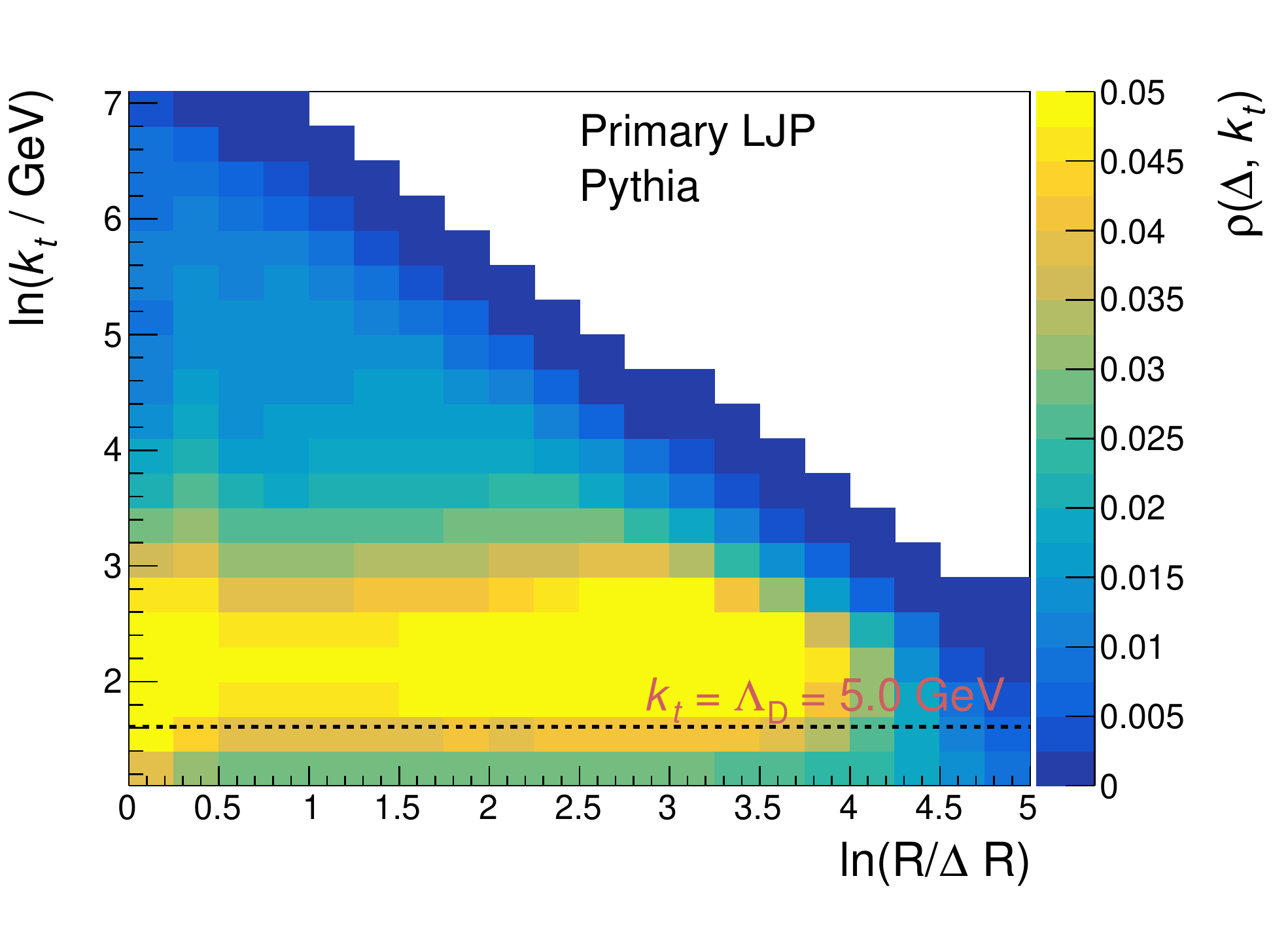}
  \includegraphics[width=0.32\textwidth]{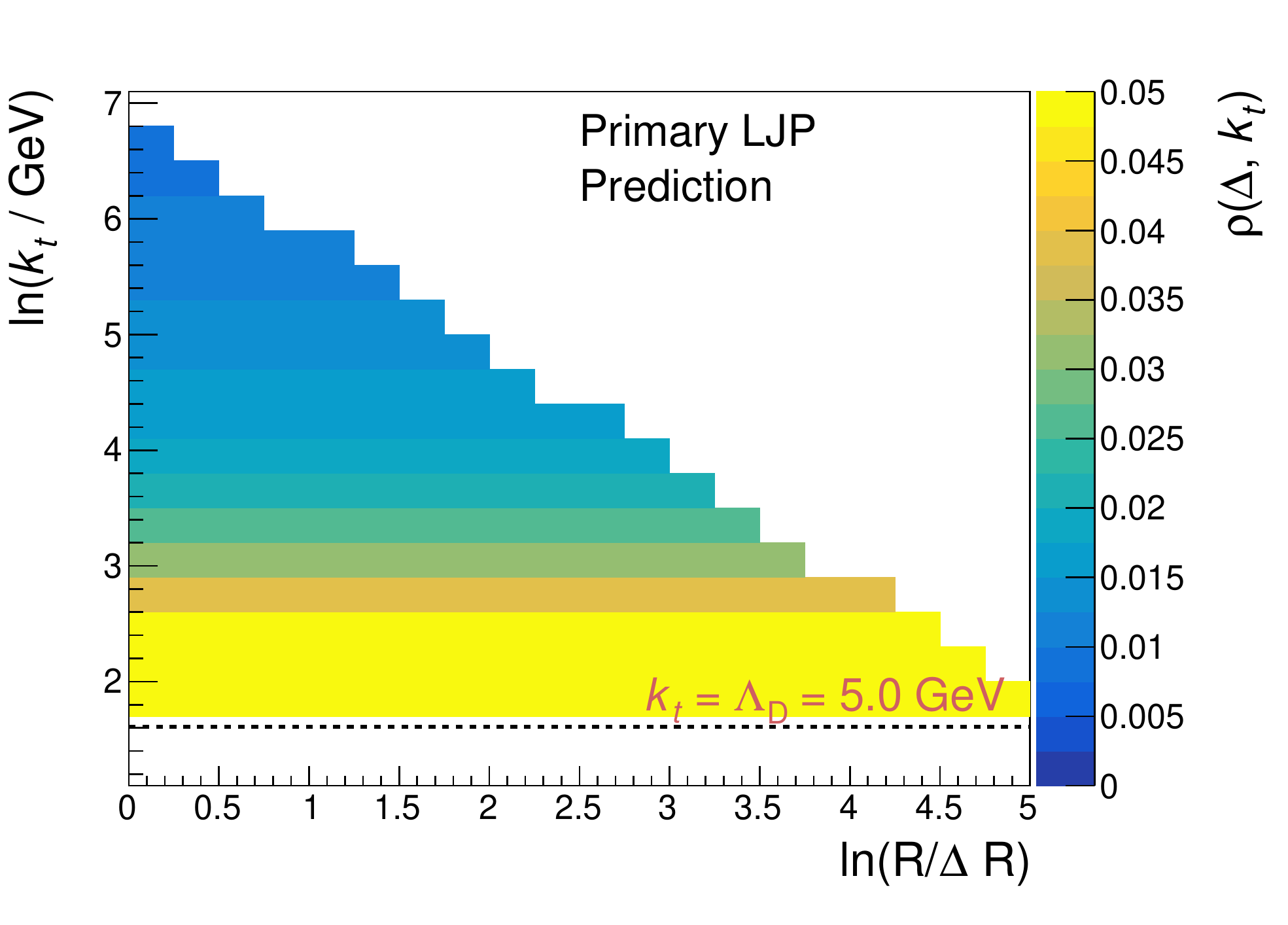}
  \includegraphics[width=0.32\textwidth]{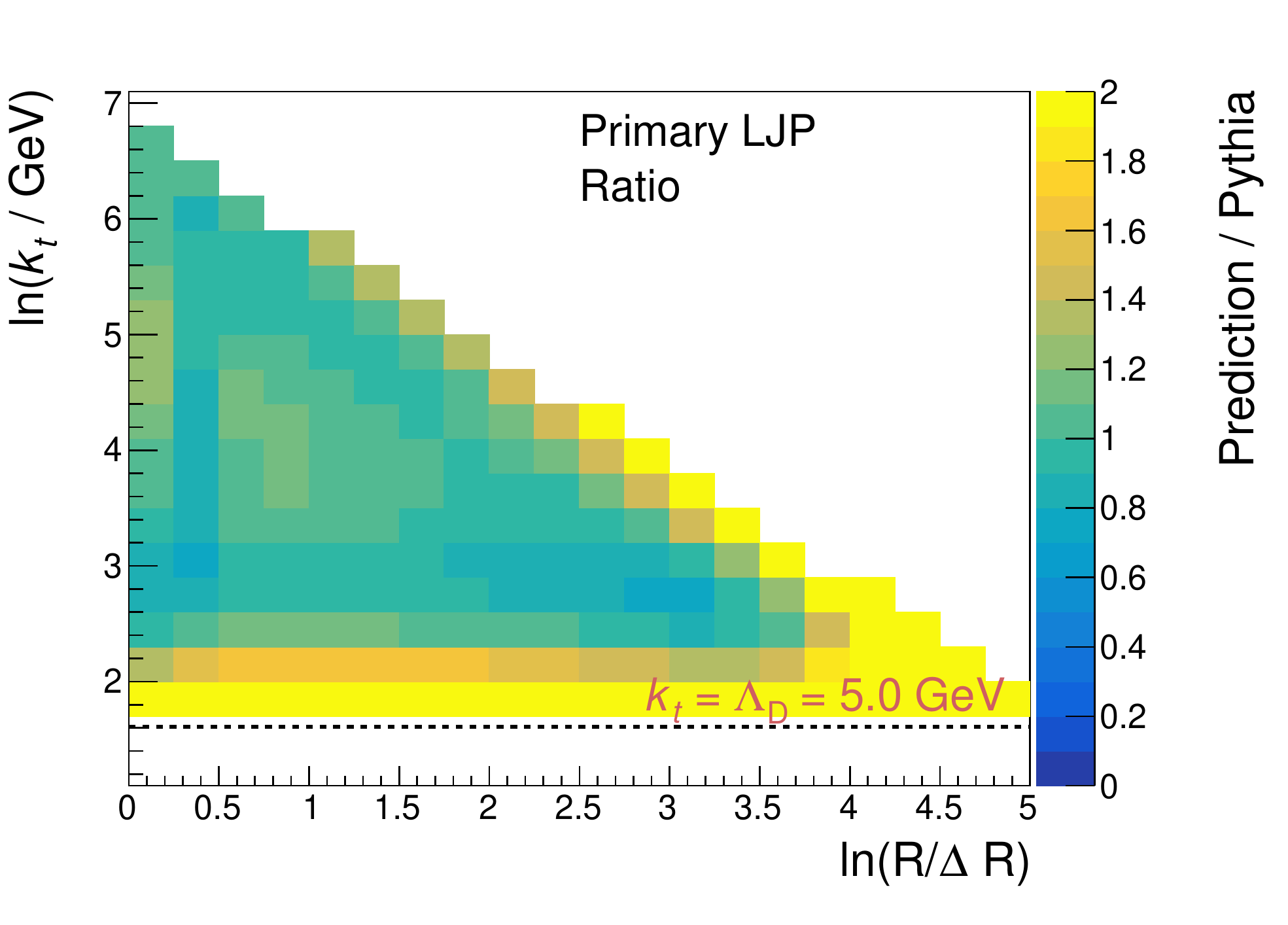}
  \caption{The LJP from dark sector quark pair production events after (top) the dark parton shower, and (middle) using the leading log prediction, and (bottom) the ratio of these.}
  \label{fig:LJP_showerOnly}
\end{figure}

\noindent 
As was explained in the original paper on the LJP~\cite{Dreyer:2018nbf}, the LJP separates into a region that is dominated by the perturbative evolution of the parton shower and a non-perturbative regime that is dominated by hadronization effects.  In the perturbative region, the leading log prediction for the density is that 
\begin{align}
\text{leading log density} \propto \frac{2\s \alpha_\text{D}(\kt)\s C_F}{\pi},
\label{eq:LL}
\end{align}
where $\alpha_\text{D}$ is the running dark gauge coupling, and $C_F = \frac{N^2-1}{2N}$ for an $SU(N)$ gauge group.  This results in increased emissions at low $\kt$ due to the running of $\alpha_\text{D}$. Higher values of $\lambdaD$ will therefore result an increasingly smaller portion of the LJP being insensitive to non-perturbative effects.  To illustrate the dark sector LJP, we first consider the LJP after the dark parton shower, but before hadronization and subsequent decays to Standard Model particles. The LJP after the dark shower is compared to the naive prediction for the LJP based on \cref{eq:LL} in \cref{fig:LJP_showerOnly}. In general, this prediction provides a good description of the behavior of the LJP for $\kt \gtrsim \lambdaD$.

\begin{figure}[thb]
  \centering
  \includegraphics[width=0.23\textwidth]{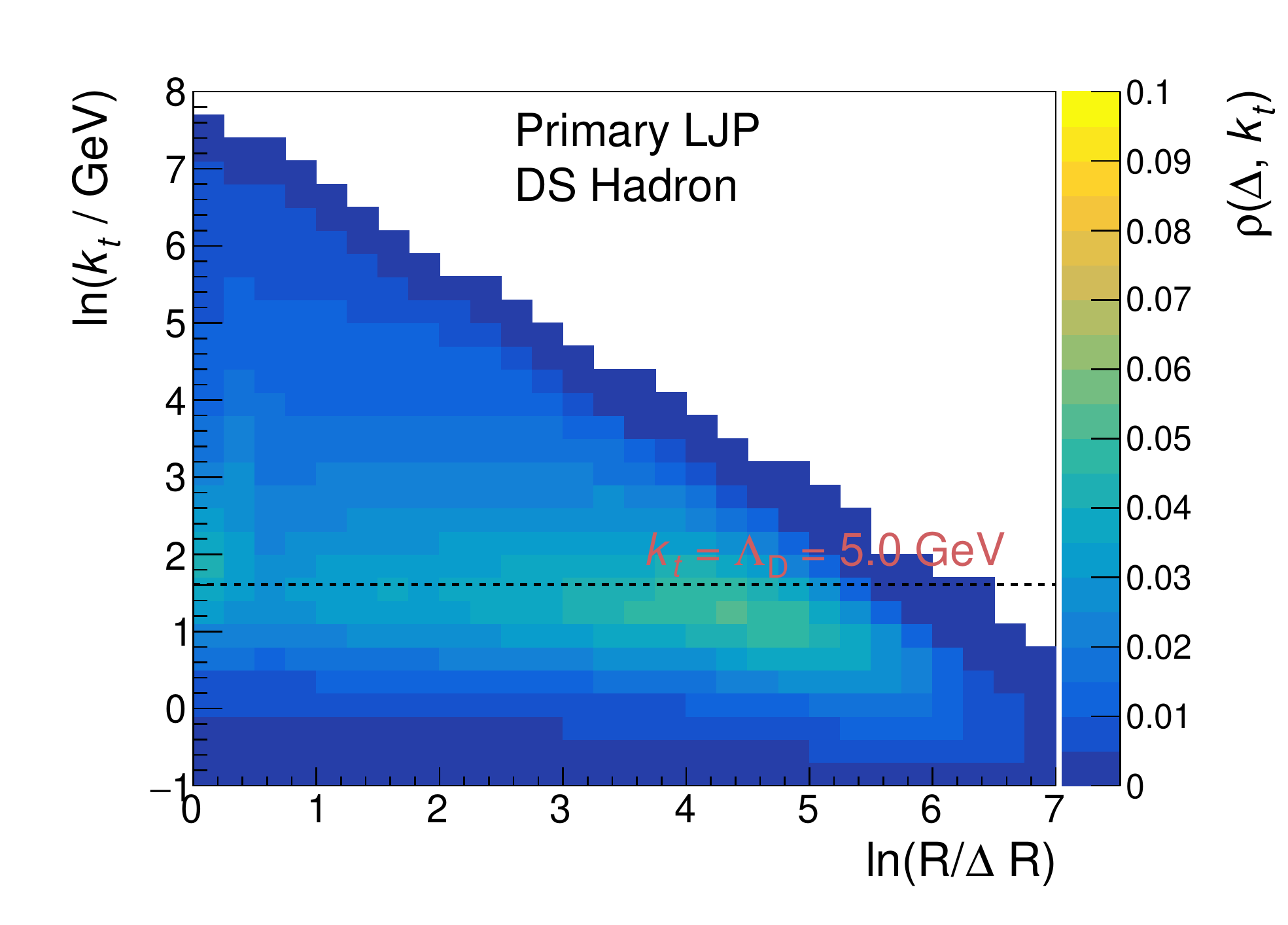}
  \includegraphics[width=0.23\textwidth]{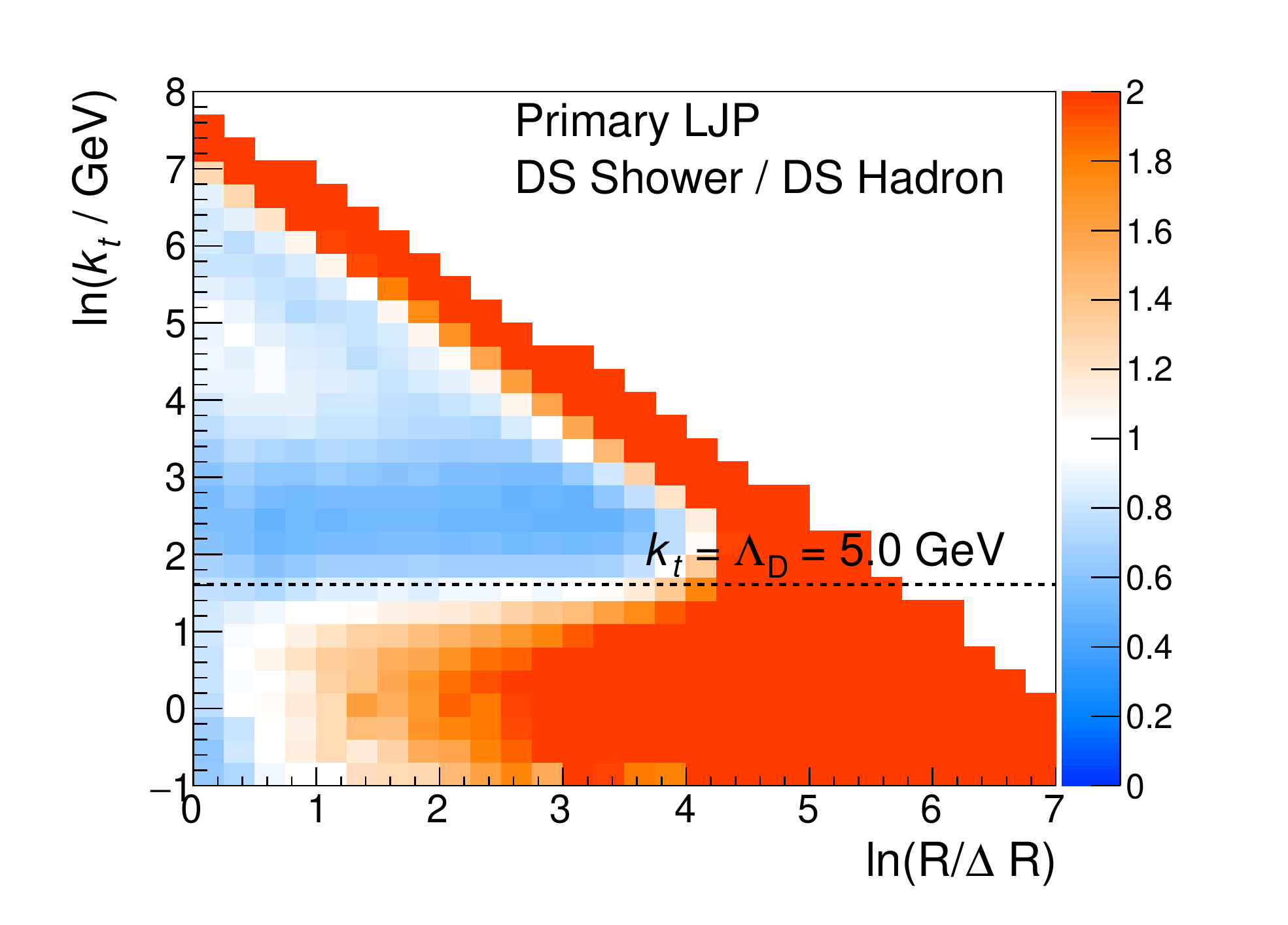} \\
  \includegraphics[width=0.23\textwidth]{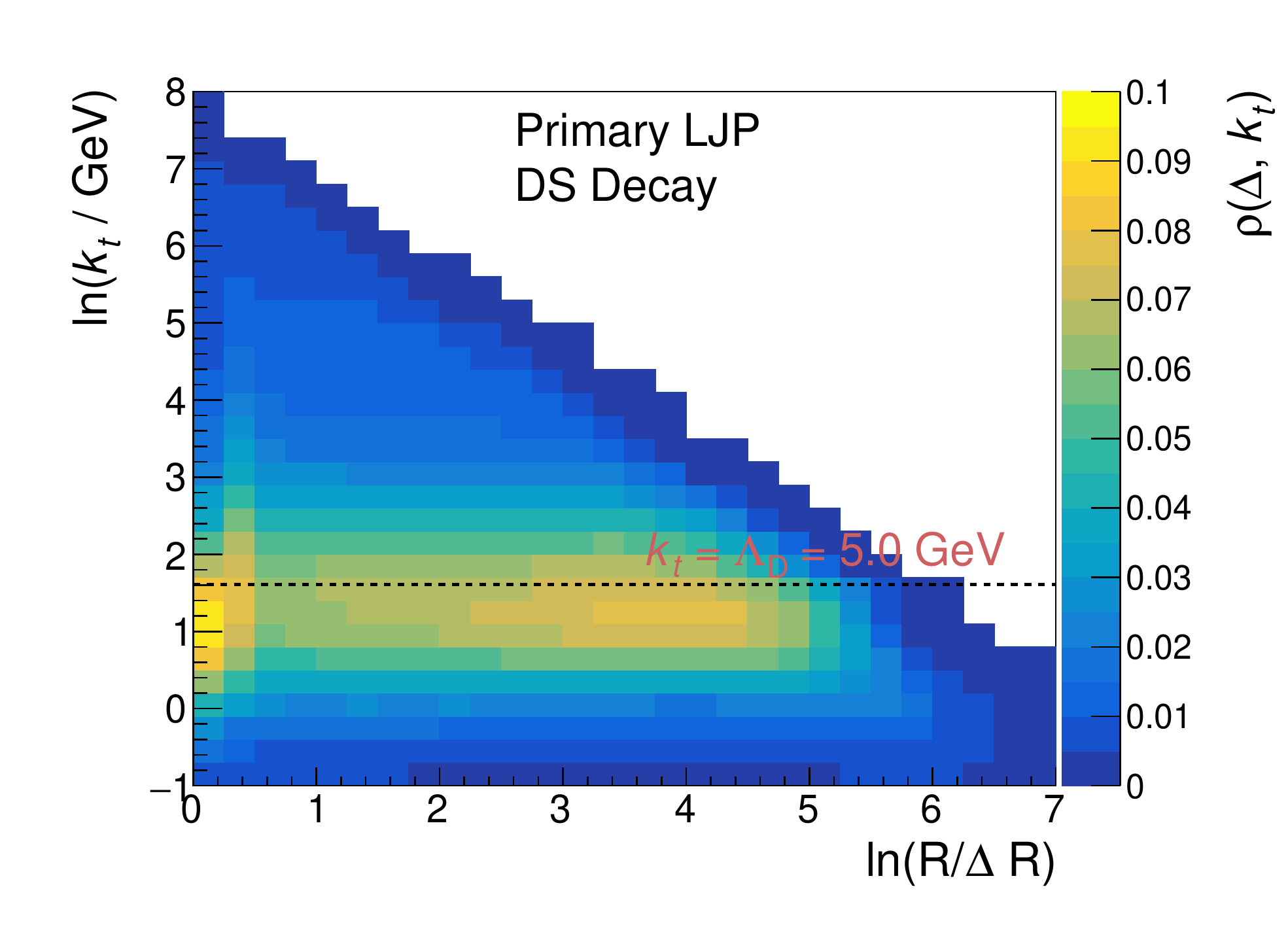}
  \includegraphics[width=0.23\textwidth]{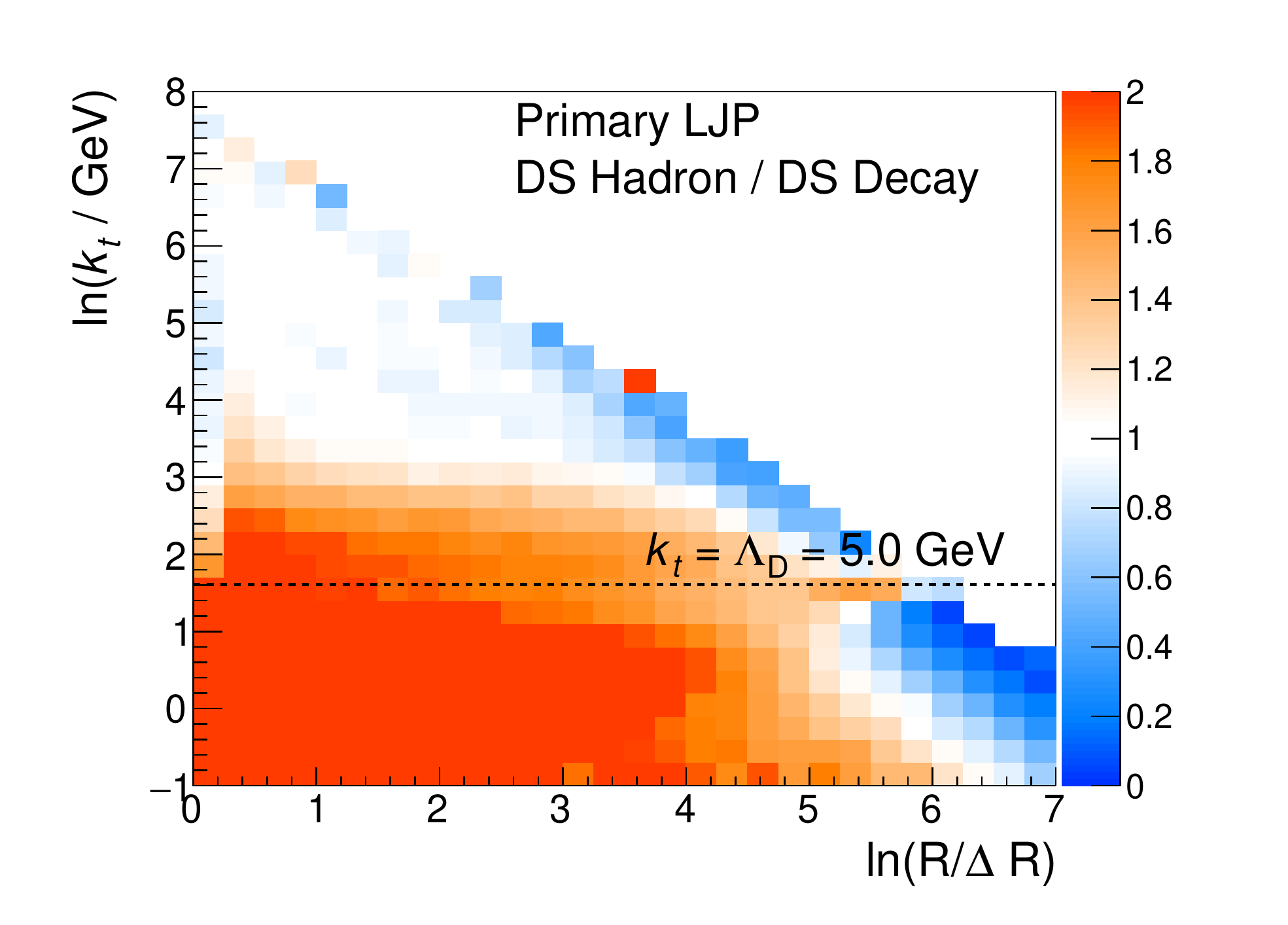} \\
  \includegraphics[width=0.23\textwidth]{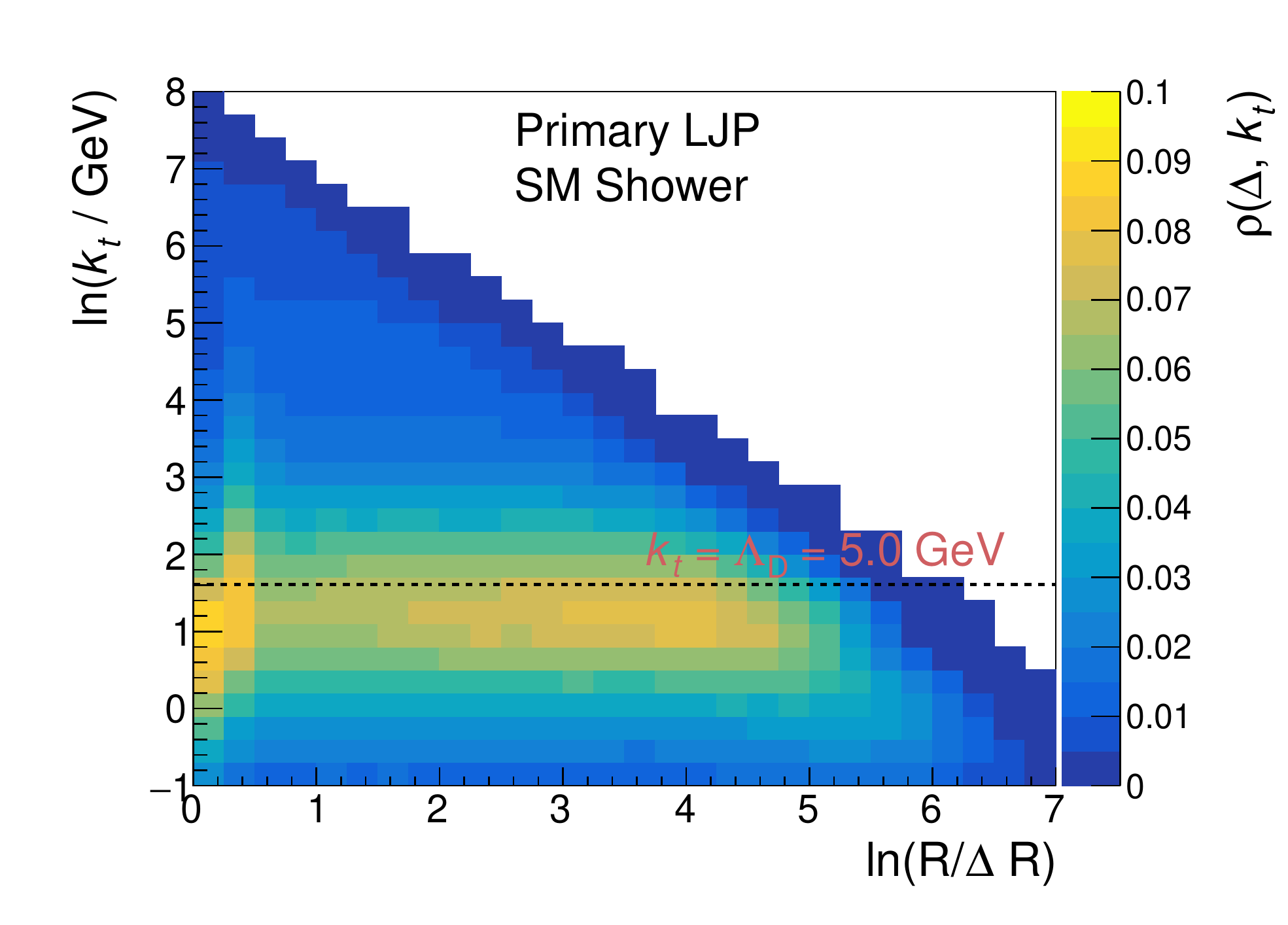}
  \includegraphics[width=0.23\textwidth]{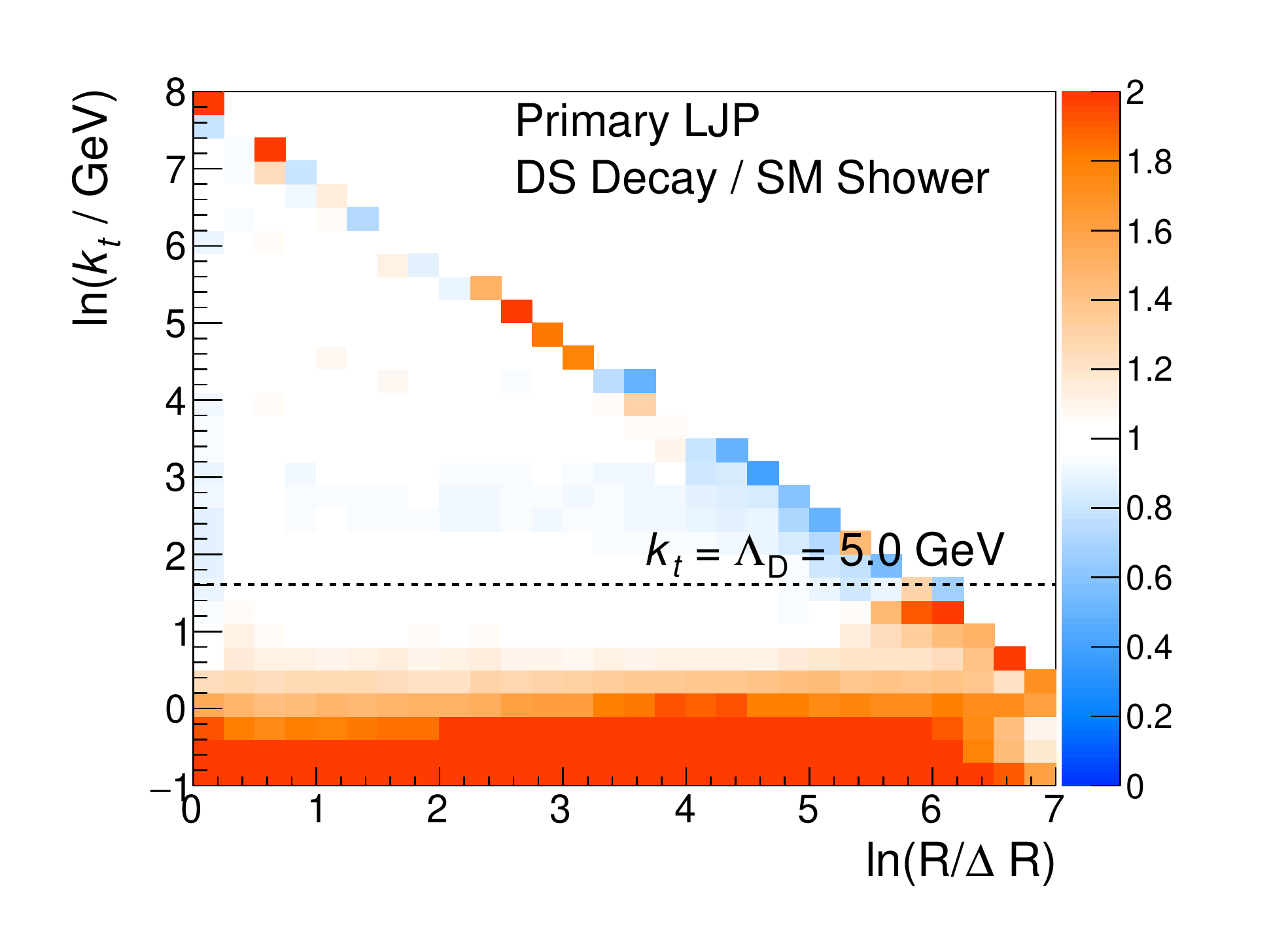} \\
  \includegraphics[width=0.23\textwidth]{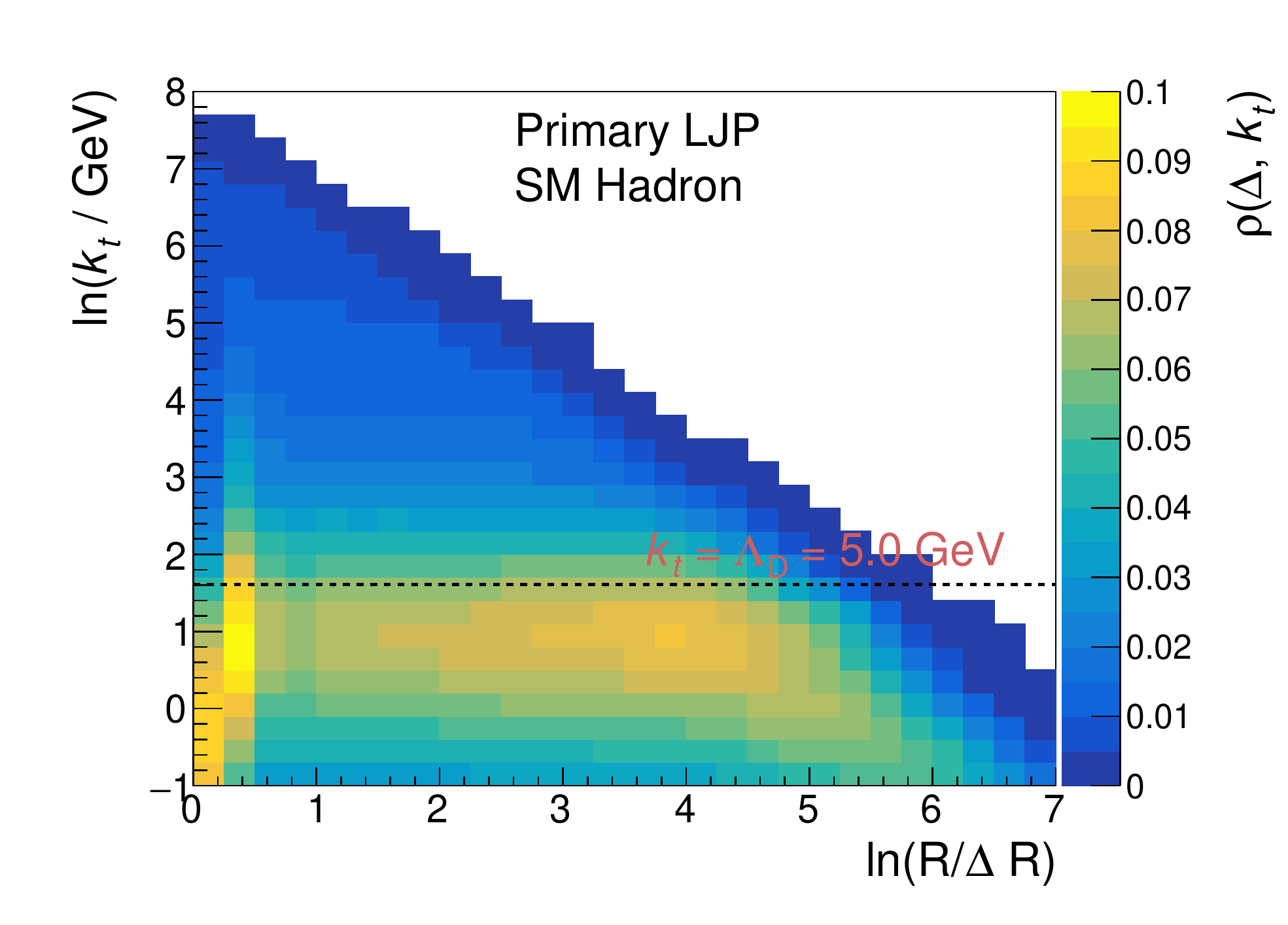} 
  \includegraphics[width=0.23\textwidth]{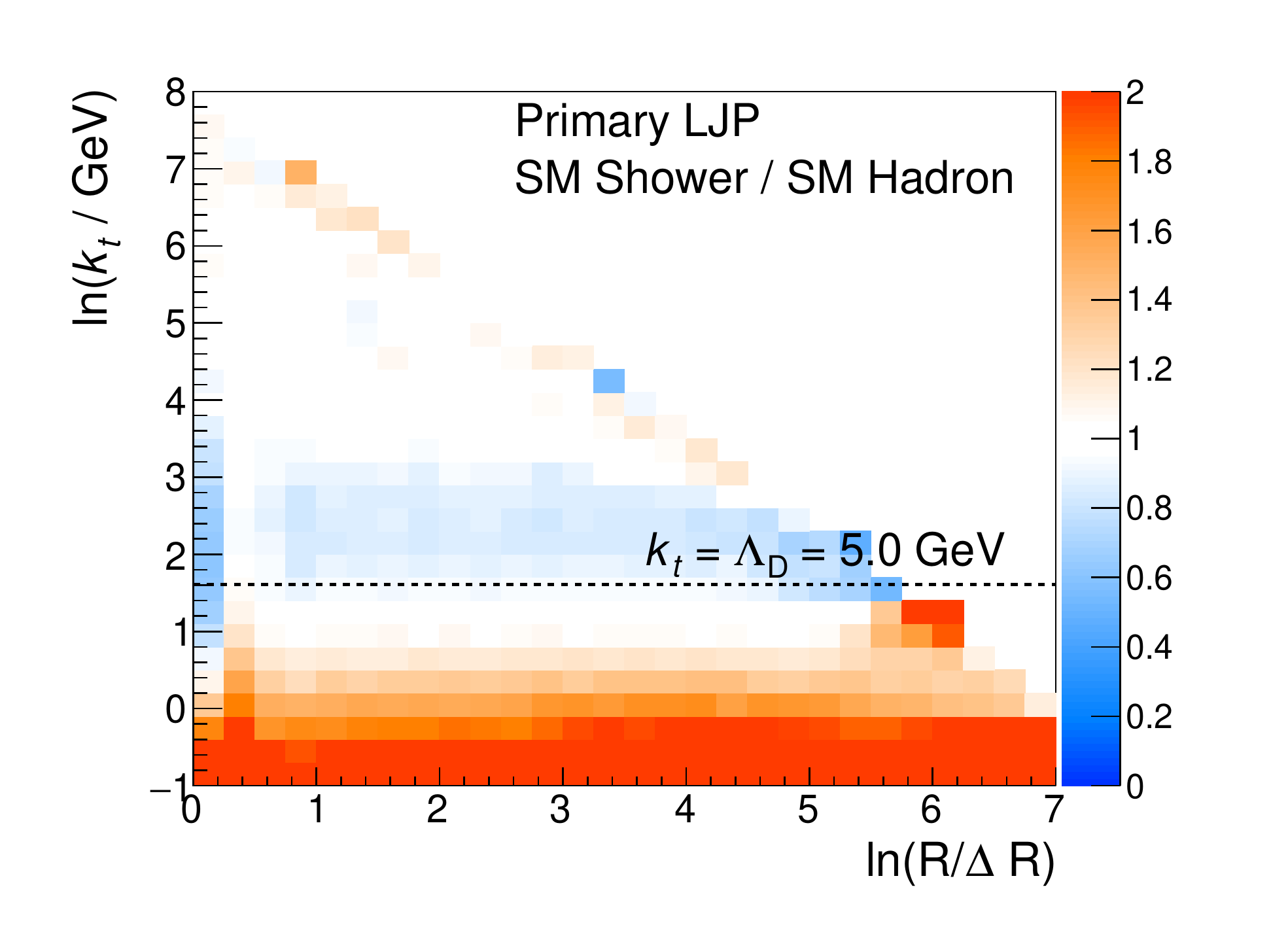} \\
  \caption{The left column provides the LJP for the following stages: (top) dark sector hadronization, (second) dark sector decay, (third) Standard Model shower, and (bottom) Standard Model hadronization.  The right column provides the ratio of the LJP for the following: (top) the dark shower to the dark hadron, (second) dark hadron to dark hadron decay, (third) dark hadron decay to Standard Model shower, and (bottom) the Standard Model shower to Standard Model hadron.}
  \label{fig:LJP}
\end{figure}

Unlike in the case of the Standard Model, the dark sector can involve widely separated scales since the dark hadron masses and the dark confinement scale are free parameters.  This can lead to additional structures in the LJP, 
which has the largest non-trivial impact at the dark sector decay stage.  Note that for our main benchmark example, the dark sector masses are taken to be of order $\lambdaD$, so these two effects lie on top of each other in the LJP.  Examples where these scales are not equal are provided in the supplemental material.
The effects of the remaining stages of event generation are shown in \cref{fig:LJP}, which provides the LJP, as well as the ratio of LJP between successive stages of the event generation. The LJP is normalized to the number of jets, such that the integral of the plane provides the average number of emissions per jet. Just as in the Standard Model LJP, we expect hadronization effects to be important for regions around the \lambdaD~scales. For the choices of parameters here, the Standard Model particles produced from the decays of the dark sector hadrons have relatively low \pt\!\!, due to the relatively low mass of the dark hadrons.  Therefore, the Standard Model parton shower has very little impact on the LJP, since they are already near the scale at which they will hadronize.

\begin{figure}[thb]
  \centering
  \includegraphics[width=0.32\textwidth]{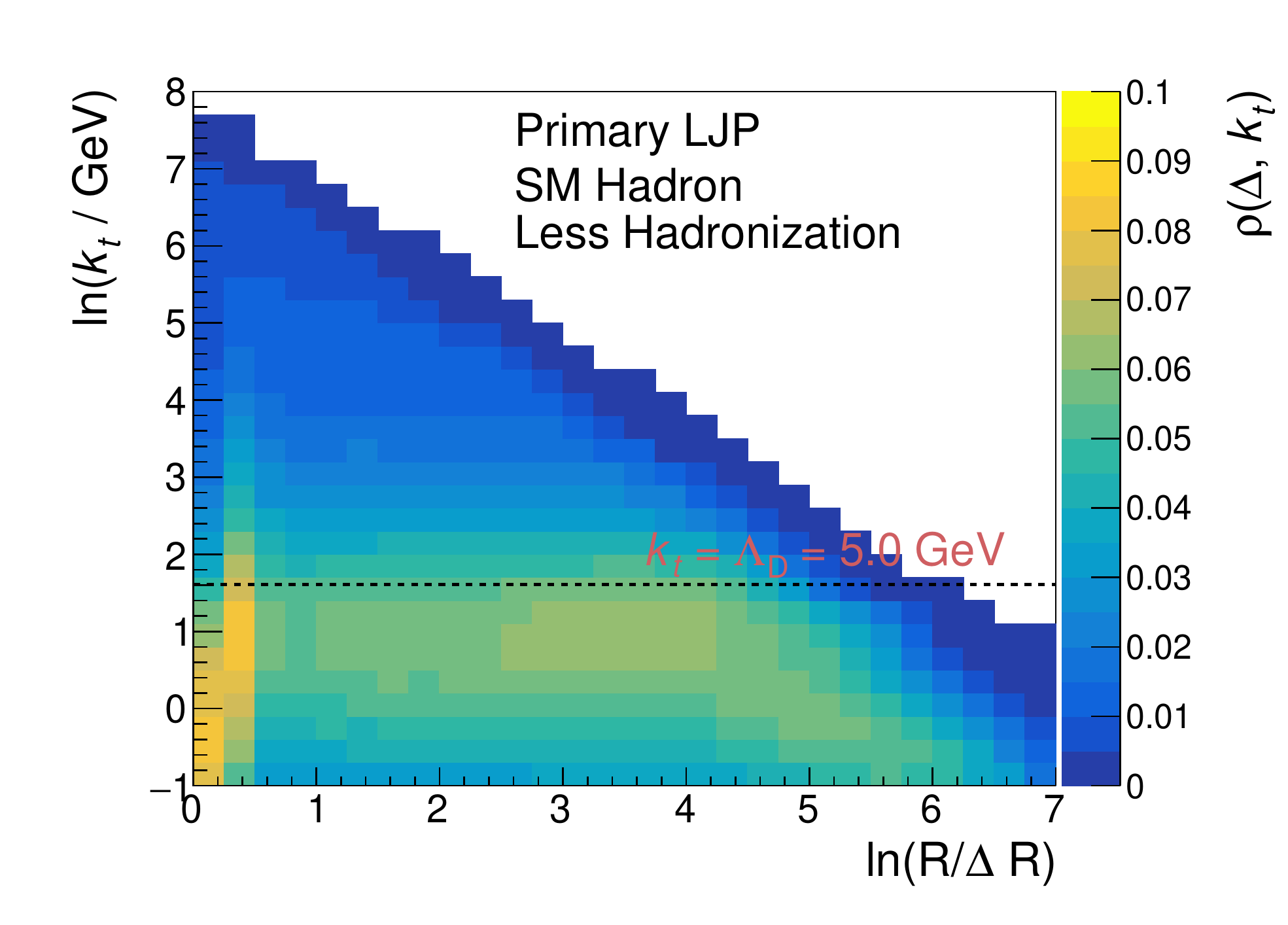}
  \includegraphics[width=0.32\textwidth]{plots/Kt_primaryBenchmarkSMHadron.pdf}
  \includegraphics[width=0.32\textwidth]{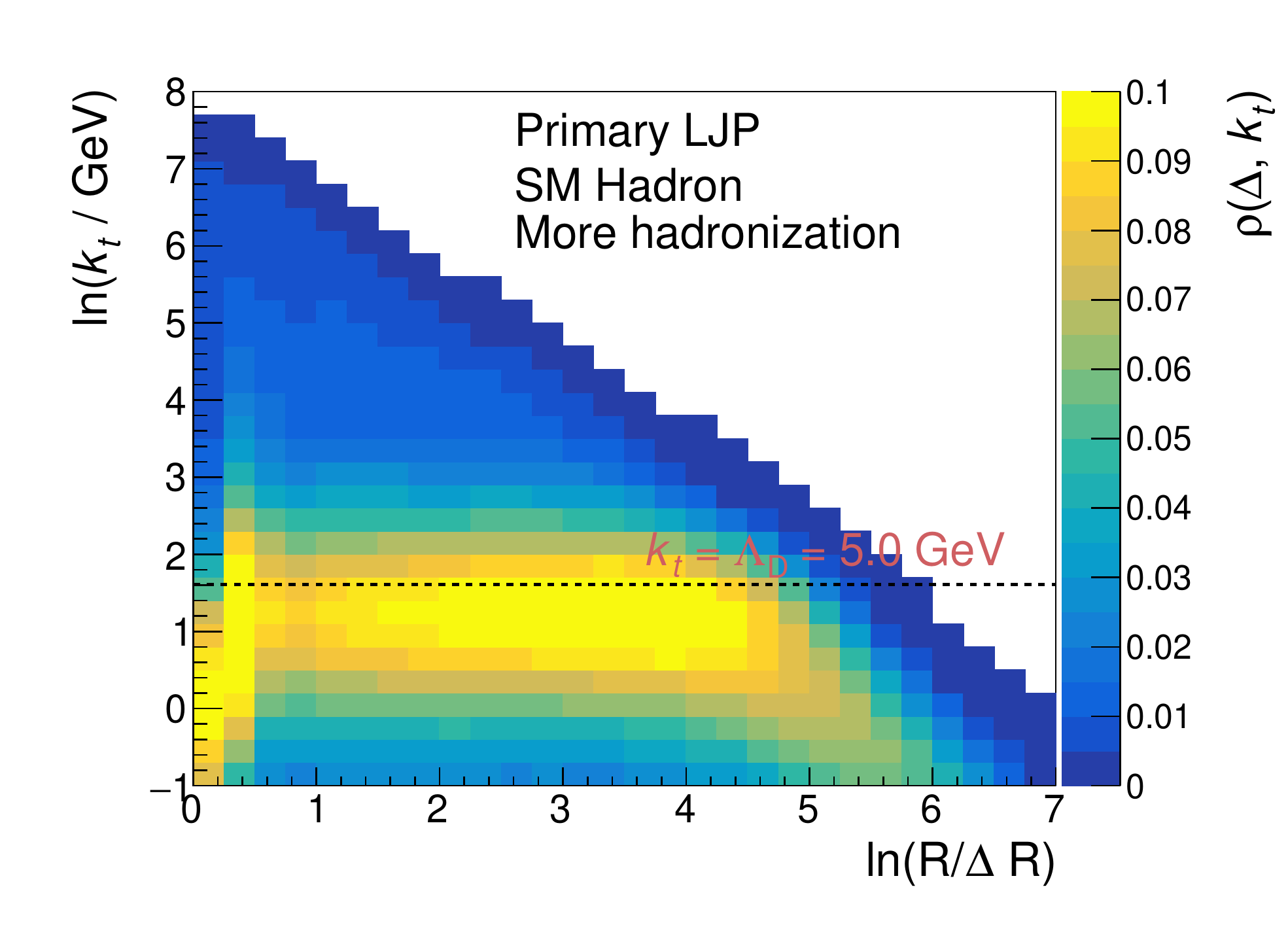}
  \caption{The LJP computed for the final state Standard Model hadrons for three different dark hadronization choices.}
  \label{fig:LJP:ld20}
\end{figure}

As demonstrated above, the behavior of the LJP for dark parton showers can be predicted at leading logarithmic accuracy, and improvements to parton shower models will enable more accurate predictions for the dark shower behavior. However, hadronization effects remain largely unconstrained, and any predictions are heavily reliant on specific models of these effects. For the Standard Model, hadronization parameters are tuned using data, but this is not possible for dark showers, nor is it obvious that the parameters used for the Standard Model are applicable to a generic dark sector. This poses significant challenges for any analysis sensitive to hadronization.

To study the impact of hadronization on different substructure observables, we produced three different MC samples with different configurations of the hadronization parameters. In the default setting, the hadronization settings are chosen to match the Monash tune~\cite{Skands:2014pea}. Then, each parameter is allowed to vary within the range allowed by the Monash tune, and the values are chosen to minimize and maximize the number of hadrons produced.  The specific parameters are given in \cref{tab:Rinv} in the supplemental material.

The results of this study are presented in \cref{fig:LJP:ld20}, which compares the LJP after the dark hadronization stage using the hadronization parameters tuned to produce the fewest hadrons, the default option used in the rest of these studies, and the parameters that yield the most hadrons. These choices result in significant differences in the LJP near \lambdaD, where non-perturbative effects are expected to be large. In the more perturbative region, the behavior of all three models is similar, since this is the region governed by the dark parton shower. This illustrates the importance of factorization, which enables the hadronization effects to be isolated to a particular region of the LJP, instead of impacting the entire distribution.
We conclude that the LJP provides a quantitative tool to isolate the unknown physics associated with dark sector hadronization.

\section{Resilience}
\label{robust}
\noindent
The ultimate goal is to apply these tools to search for the signatures of dark sector showers at the LHC.  To demonstrate that the LJP provides a useful tool, we will evaluate two different metrics: the background rejection and the resilience against non-perturbative effects. For a specific cut on an generic observable, the background rejection can be quantified as
\begin{equation}
p = \frac{\epsilon_{\text{D}}}{\sqrt{\epsilon_\text{QCD}}},
\end{equation}
where $\epsilon_{\text{D}}$ is the dark sector jet efficiency, and $\epsilon_\text{QCD}$ is the efficiency for Standard Model jets. In both cases, these are determined after Standard Model hadronization. The resilience against non-perturbative effects, $\zeta$, can be quantified using the impact of the hadronization on the tagger performance. In particular, we define
\begin{equation}
\zeta = \left(\frac{\Delta \epsilon_{\text{D}}}{\langle\epsilon_{\text{D}}\rangle}\right)^2,
\end{equation}
where $\Delta \epsilon = \epsilon - \epsilon'$ and $\langle\epsilon\rangle = (\epsilon + \epsilon')/2$,
For these studies, $\epsilon$ is determined using the default hadronization parameters, while $\epsilon'$ is determined using the hadronization parameters that produce the maximum number of hadrons.

As a simple example of the impact of hadronization modeling on the regions of the LJP that we are sensitive to, we consider the resilience of a tagger that counts that number of emissions in the primary LJP above a certain \kt~cut, as well as a comparison to three other observables: the jet energy sharing $D_{p_{t}}=\sqrt{\sum_{i} p_{T,i}^2}/\sum_{i} p_{T,i}$, the number of jet constituents $N_{\mathrm{constit}}$, and the jet mass. While it is possible to create more sensitive observables that make use of the 2-dimensional plane, the number of emissions provides a good proxy for the overall behavior, and the \kt~cut provides a proxy for controlling the amount of hadronization effects that are included. 

\begin{figure}[t!]
  \centering
  \includegraphics[width=0.4\textwidth]{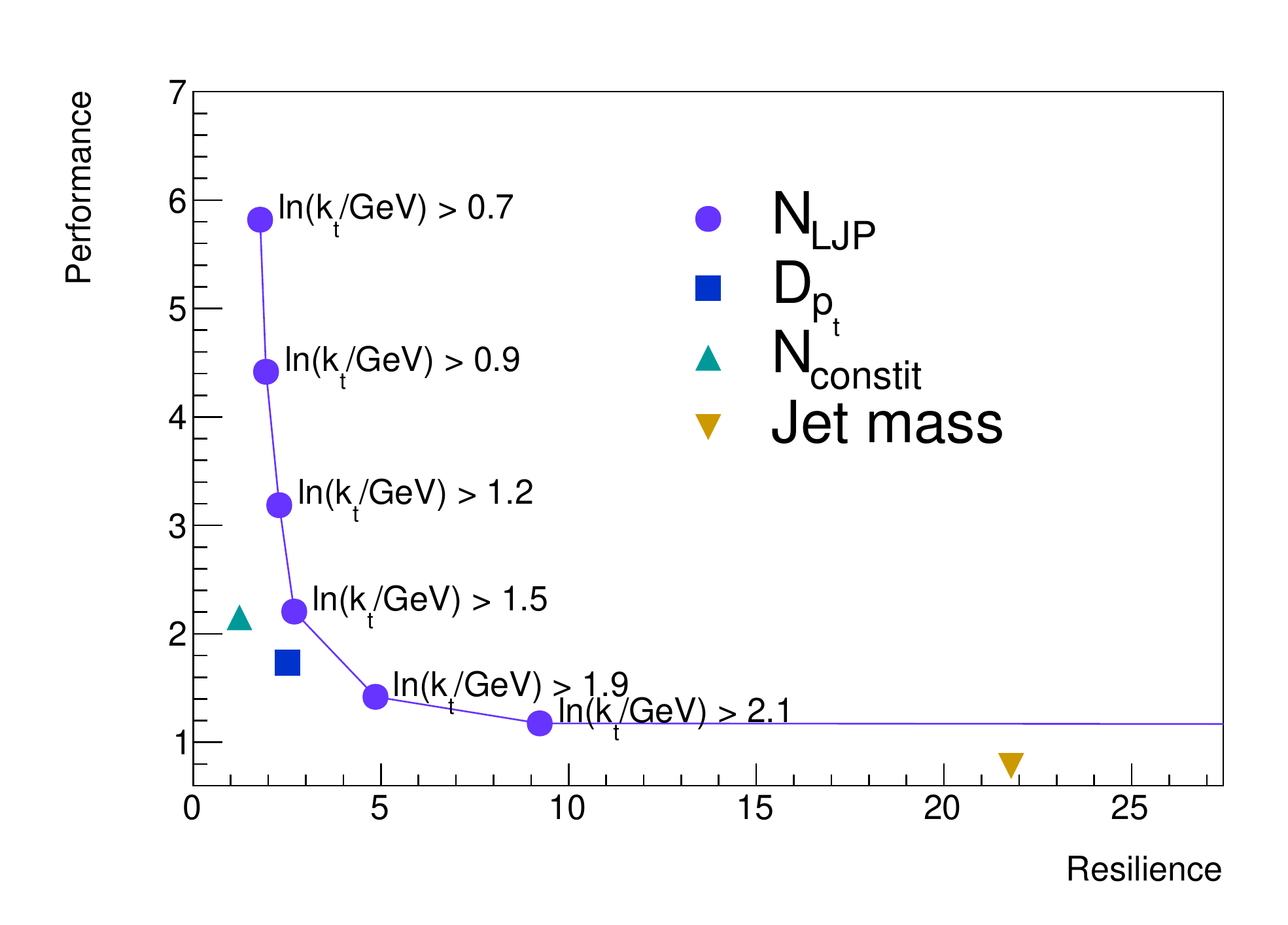}
  \caption{The performance versus resilience of different substructure cuts.  The LJP can be used as a powerful discriminator without introducing too much sensitivity to the underlying hadronization uncertainty.}
  \label{fig:nconstit}
\end{figure}

The results of this study are shown in \cref{fig:nconstit}. In general, for a given tagger performance, the LJP provides better resilience than the other observables. Even for choices which reduce the sensitivity, this comes with the benefit of greater interpretability. Even if the hadronization modeling is incorrect, this observable can still be used to set robust limits on dark sector models. With observables like the number of constituents, where the tagging performance is greatly impacted by the specific hadronization parameter choices, any results would be difficult to interpret generically for hadronization variations, even with the same $n_{C}$, $n_{F}$, and \lambdaD~parameters.

\section{Conclusions and Outlook}
\label{conclusions}
\noindent
In this letter, we have proposed applying the Lund jet plane (LJP) as a tool for dark sector searches at the LHC.  One of the key benefits of the LJP is that it isolates the non-perturbative effects of hadronization.  The effects of dark sector hadronization on jet substructure observables are in principle incalculable, and one must appeal to phenomenological models.  By cutting away the region of the LJP that is most sensitive to hadronization, one can perform robust searches for these models including the effects of substructure.  

While this letter provides compelling evidence for this application of the LJP, there are many future directions to explore.  In this study, we decayed all the dark hadrons to the Standard Model.  But the general expectation is that some fraction of the dark hadrons would be stable.  It would be interesting to study the interplay of the LJP efficiency in the presence of stable dark matter candidates.  It would also be useful to test the efficacy of the LJP as a tool in the context of more realistic searches, involving constraints from other searches and so on.  Additionally, models with a seperation of scales between the dark hadronization scale and the dark meson masses can lead to new features in the LJP (see \cref{fig:lambda50:LJP} in the supplemental material) that could be exploited when designing searches. Another important role for the LJP is to compare the predictions of different dark sector simulations, as a way to quantify the variations among their predictions.  This could in principle be utilized to make these simulation tools more robust.  It would also be very interesting to apply machine learning tools to the LJP (see~\eg~\cite{Dreyer:2020brq}), which would allow to maximize the information contained in the LJP in a search for new physics.  Developing robust searches for dark sector showers is of paramount importance, and this letter shows that the LJP provides an ideal tool for this task.

\acknowledgments
\noindent
We are grateful to Joel Doss for collaboration at early stages of this work, and to Fr\'ed\'eric Dreyer and Gr\'egory Soyez for useful discussions.
The work of T.~Cohen is supported by the U.S.~Department of Energy under grant number DE-SC0011640.
The work of J.~Roloff is supported by the U.S. Department of Energy, Office of Science, Office of High Energy Physics under contract no. DE-SC0012704.
The work of C.~Scherb is supported by the  Office of High Energy Physics of the U.S. Department of Energy under Contract
No. DE-AC02-05CH11231 and the Alexander von Humboldt Foundation.

\end{spacing}

\bibliographystyle{utphys}
\bibliography{darkShowers}

\clearpage
\onecolumngrid
\section*{SUPPLEMENTAL MATERIAL}
\setcounter{section}{0}
\renewcommand{\thesection}{\Alph{section}}
\section{Benchmark Points}
\begin{table}[hb]
  \renewcommand{\arraystretch}{1.6}
  \setlength{\arrayrulewidth}{.3mm}
  \centering
  \small
  \setlength{\tabcolsep}{0.35 em}
	\begin{tabular}{l|c|c|c|c|c|c|c|c|c|c}
		&$n_C$&$n_F$&$\lambdaD$&$m_{Q_\text{D}}$&$m_{\pi_\text{D}}$&$m_{\rho_\text{D}}$&$a_L$&$b_{m_{Q_\text{D}}^2}$&$r_{Q_\text{D}}$&fraction $\rho_\text{D}$\\[3pt]
		\hline
		primary benchmark&3&3&5~GeV&2.5~GeV&5~GeV&5~GeV&0.3&0.8&1&default ($=0.75$)\\
        large hadronization&3&3&5~GeV&2.5~GeV&5~GeV&5~GeV&2&0.2&2&default\\
		small hadronization&3&3&5~GeV&2.5~GeV&5~GeV&5~GeV&0&2&0&default\\
        snowmass benchmark&3&2&6.5~GeV&0.5~GeV&10~GeV&20~GeV&0.3&0.8&1&default\\
		dark QCD scale $\lambdaD$&3&3&50~GeV&2.5~GeV&5~GeV&5~GeV&0.3&0.8&1&default\\
	\end{tabular}
	\caption{Dark sector parameters for the benchmark points used.}
	\label{tab:Rinv}
\end{table}

\section{Performance with the Snowmass benchmark}
\begin{figure}[thb]
  \centering
   \includegraphics[width=0.38\textwidth]{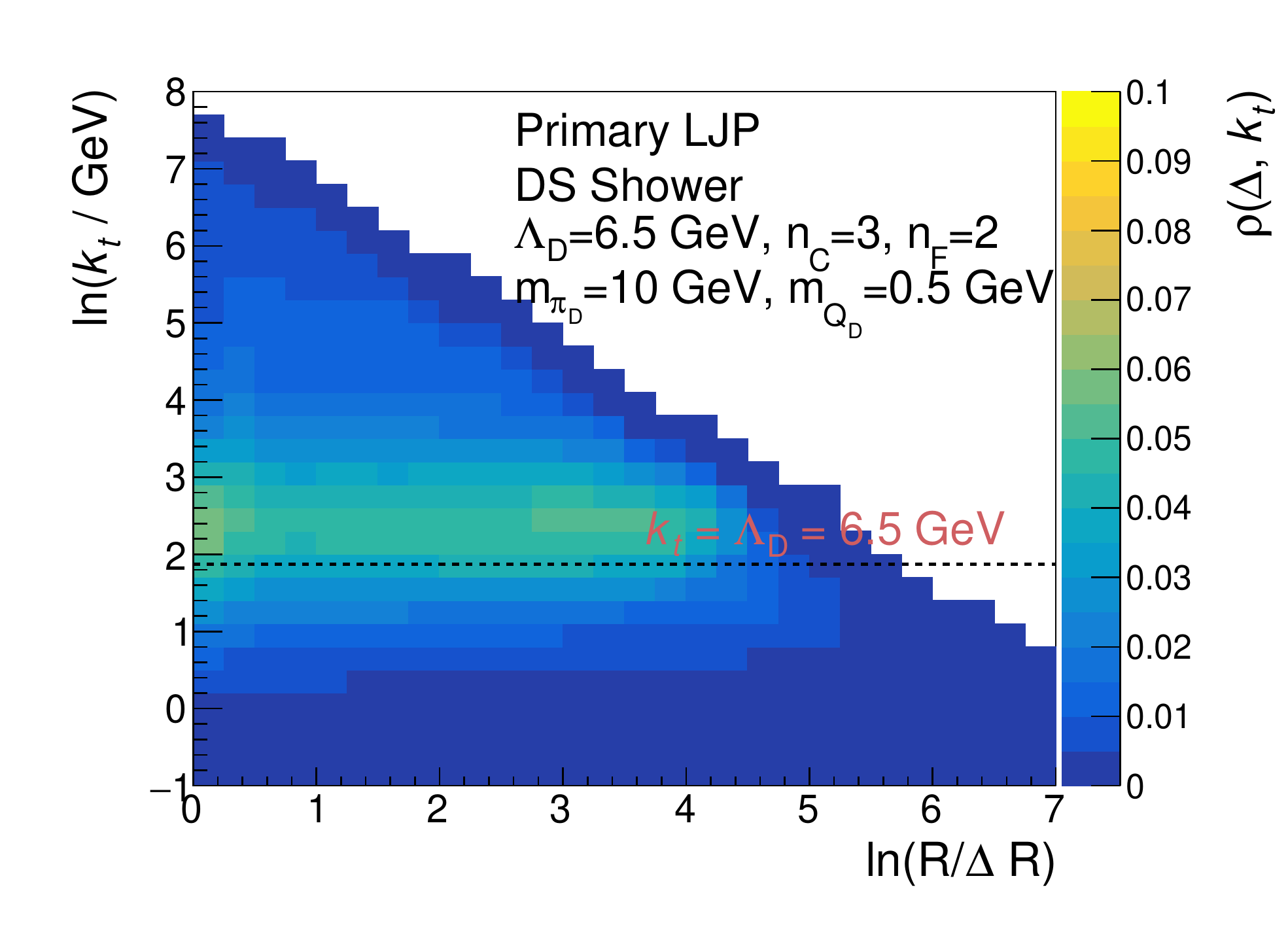}
  \leavevmode\phantom{\includegraphics[width=0.38\textwidth]{plots/snowmassBenchmarkPlots/Kt_Snowmass_DMShower.pdf}}
  \includegraphics[width=0.38\textwidth]{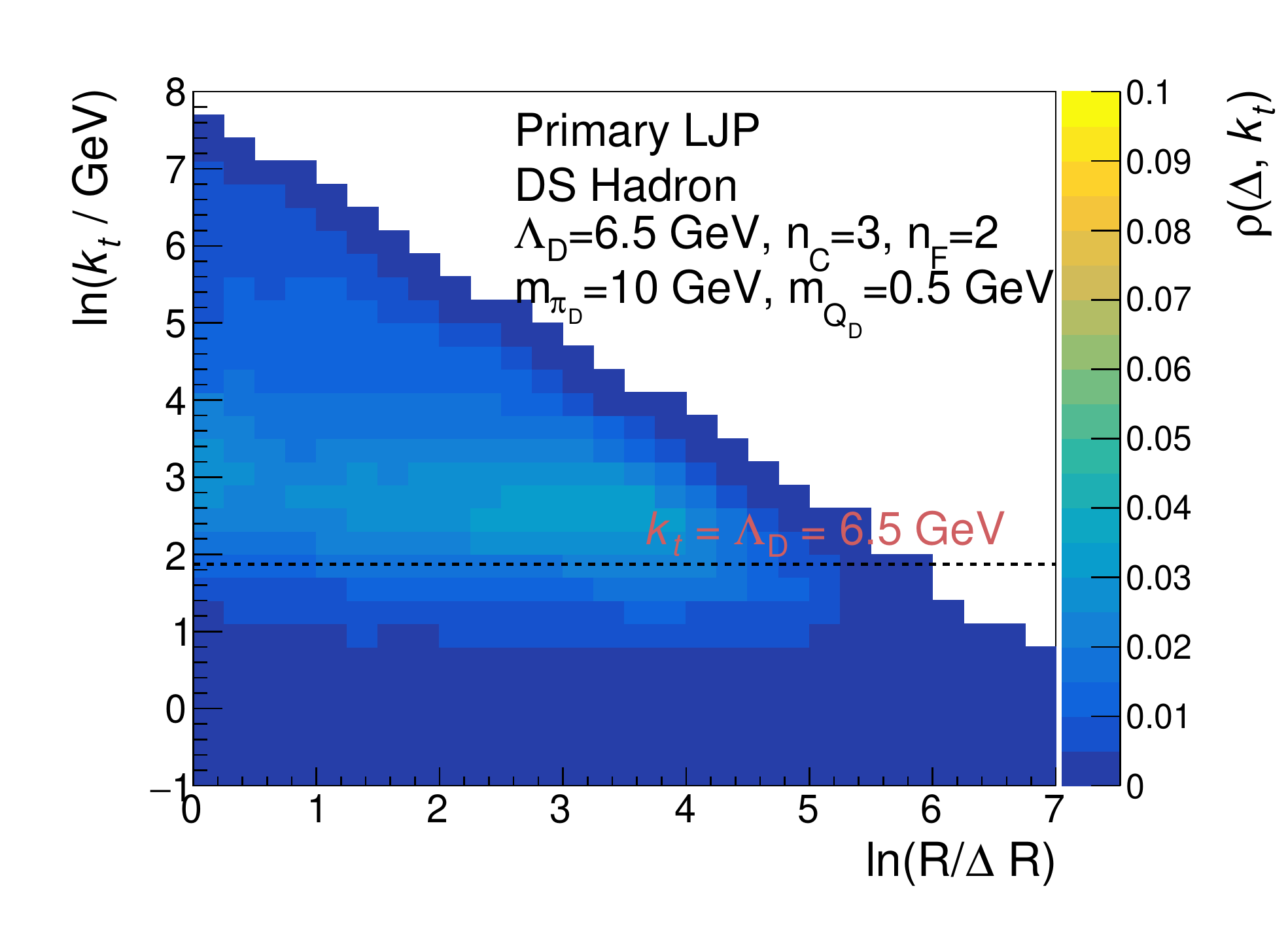}
  \includegraphics[width=0.38\textwidth]{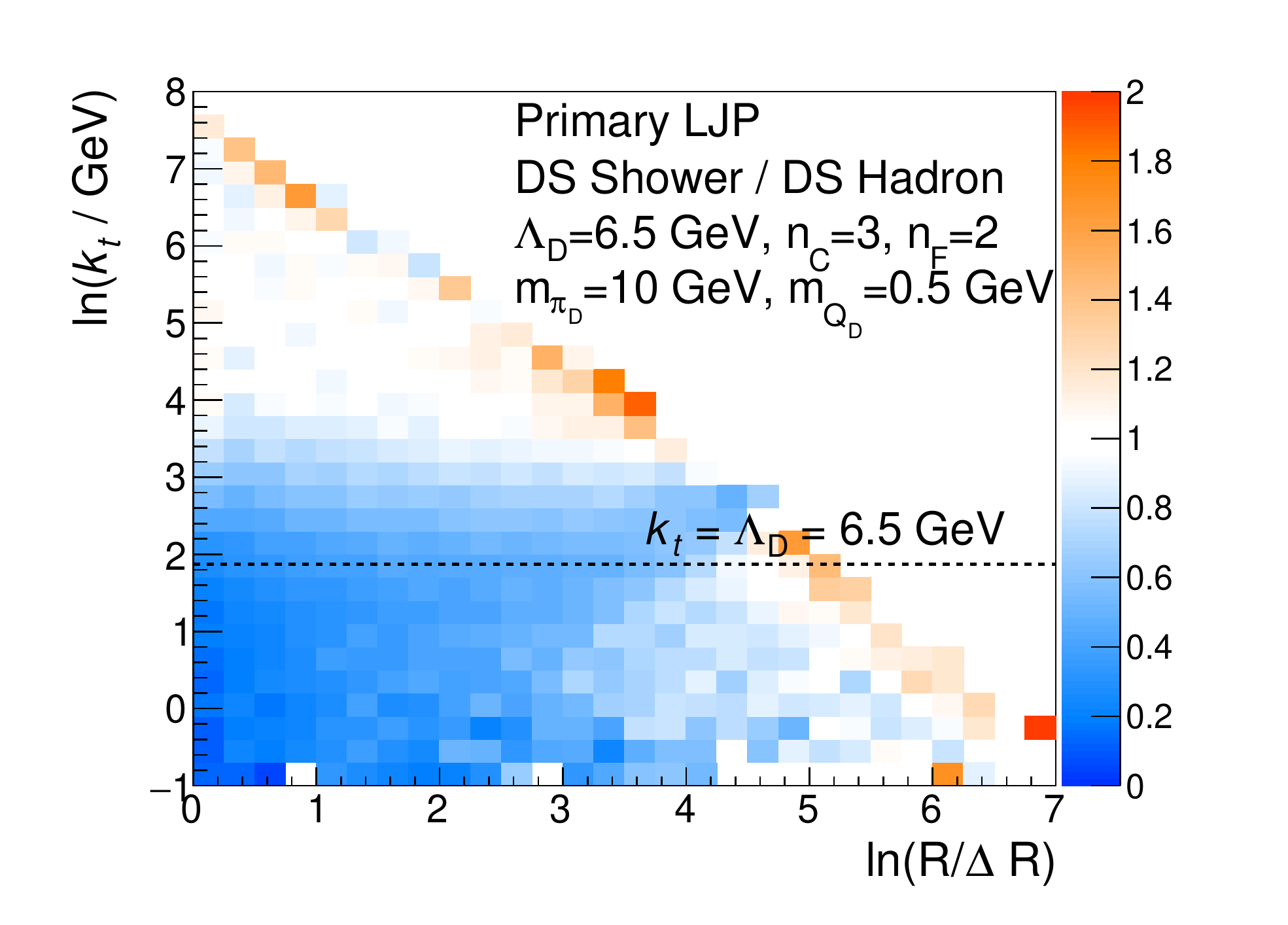} \\
  \includegraphics[width=0.38\textwidth]{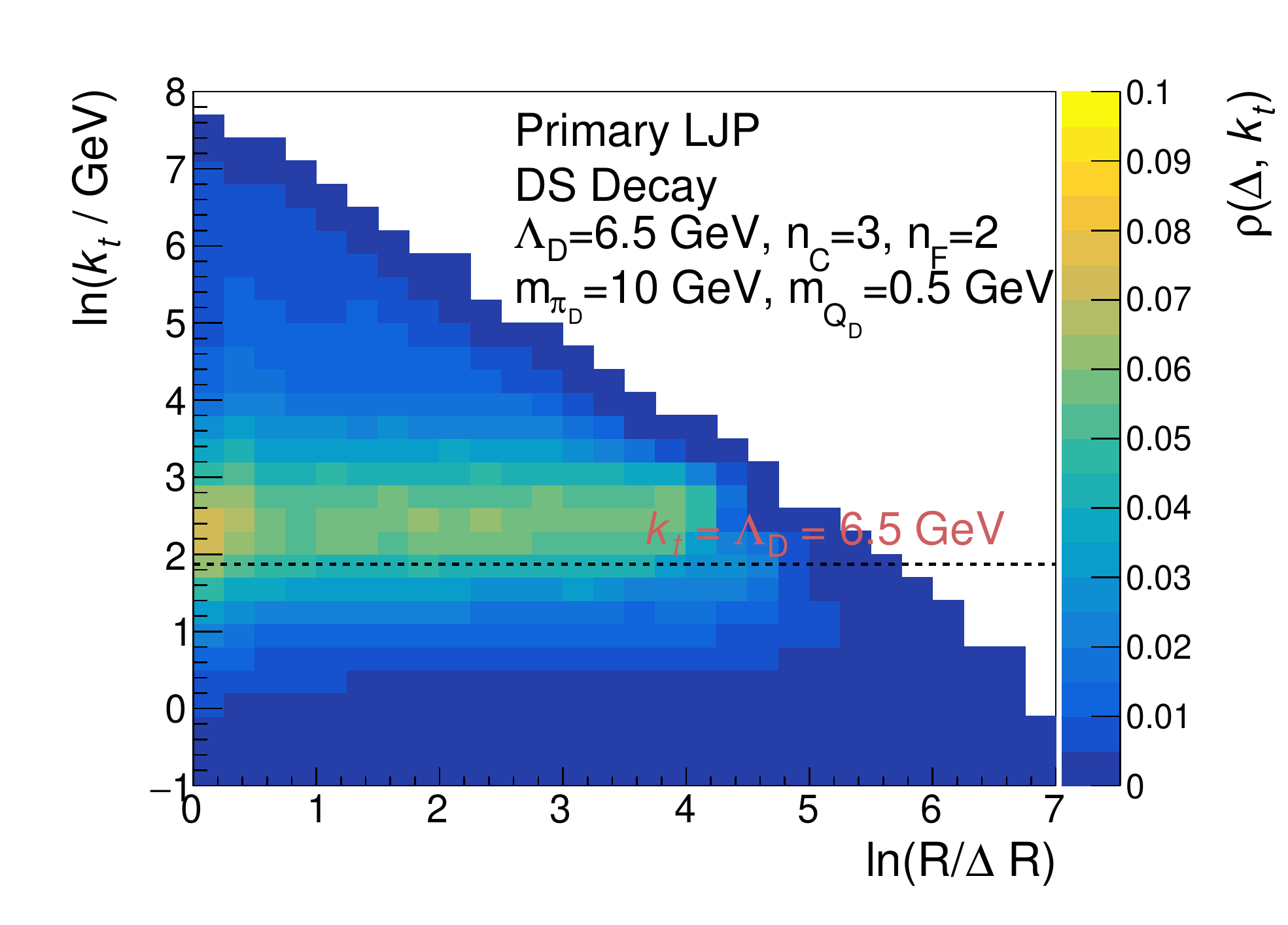}
  \includegraphics[width=0.38\textwidth]{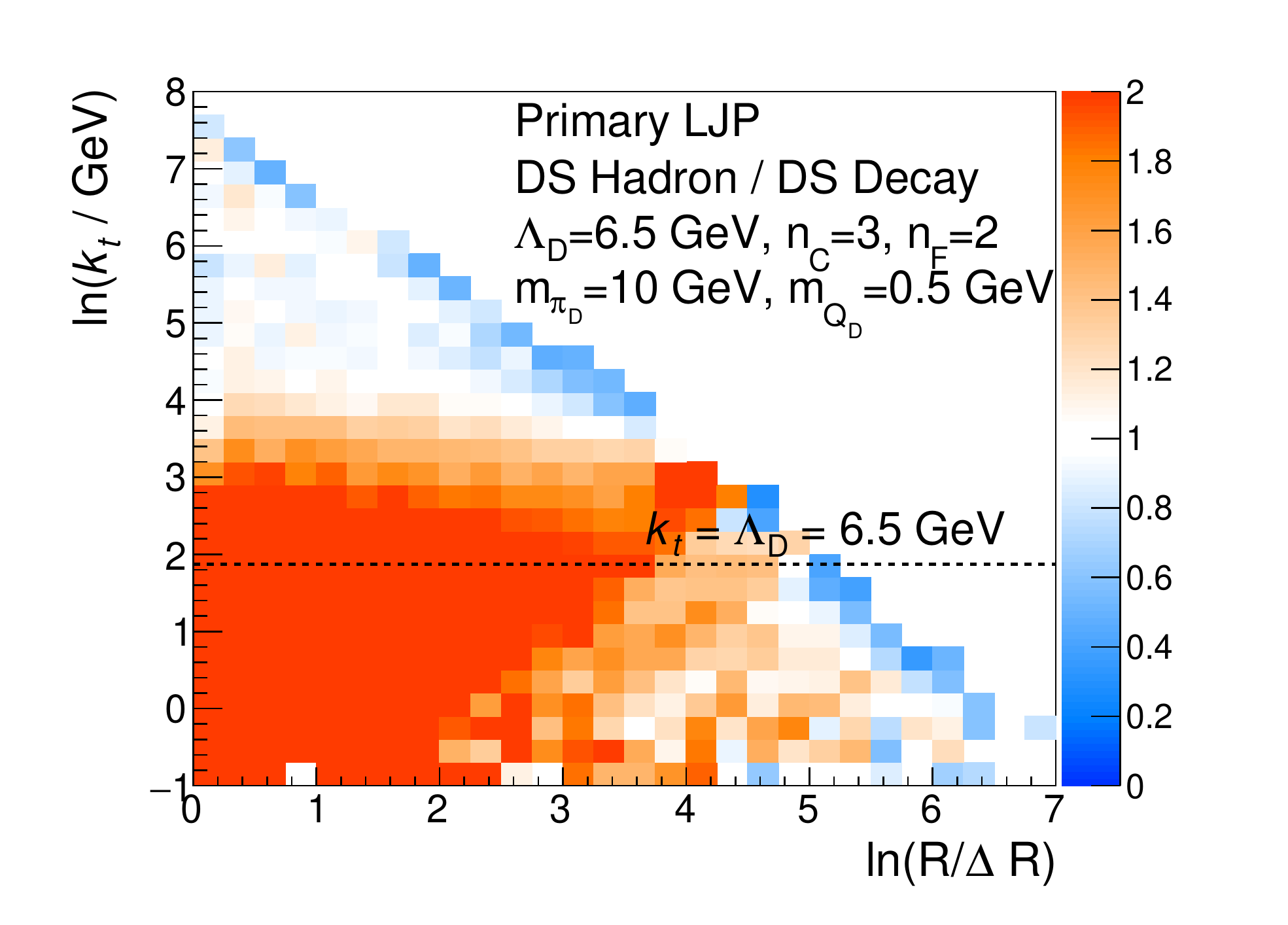} \\
  \includegraphics[width=0.38\textwidth]{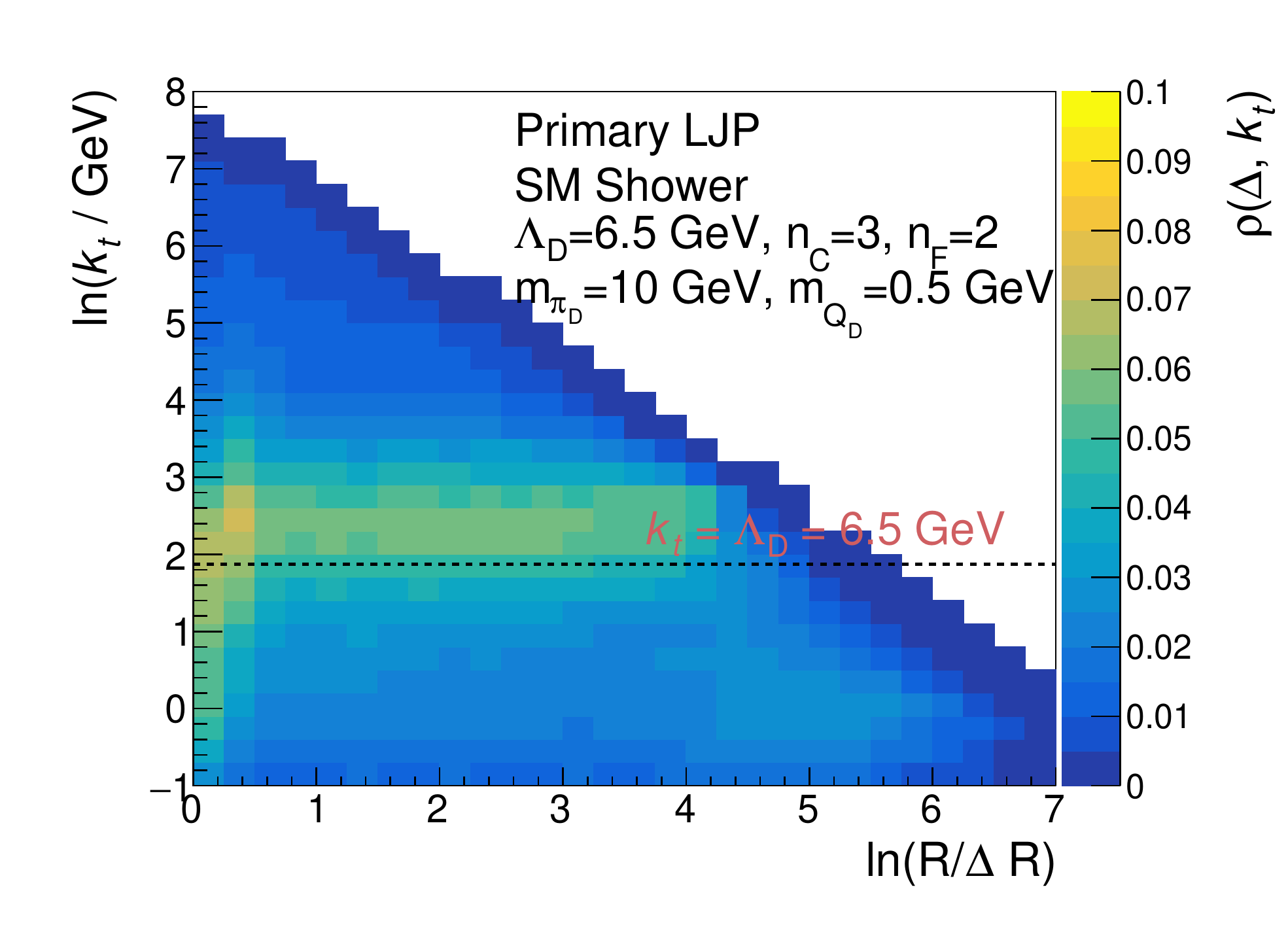}
  \includegraphics[width=0.38\textwidth]{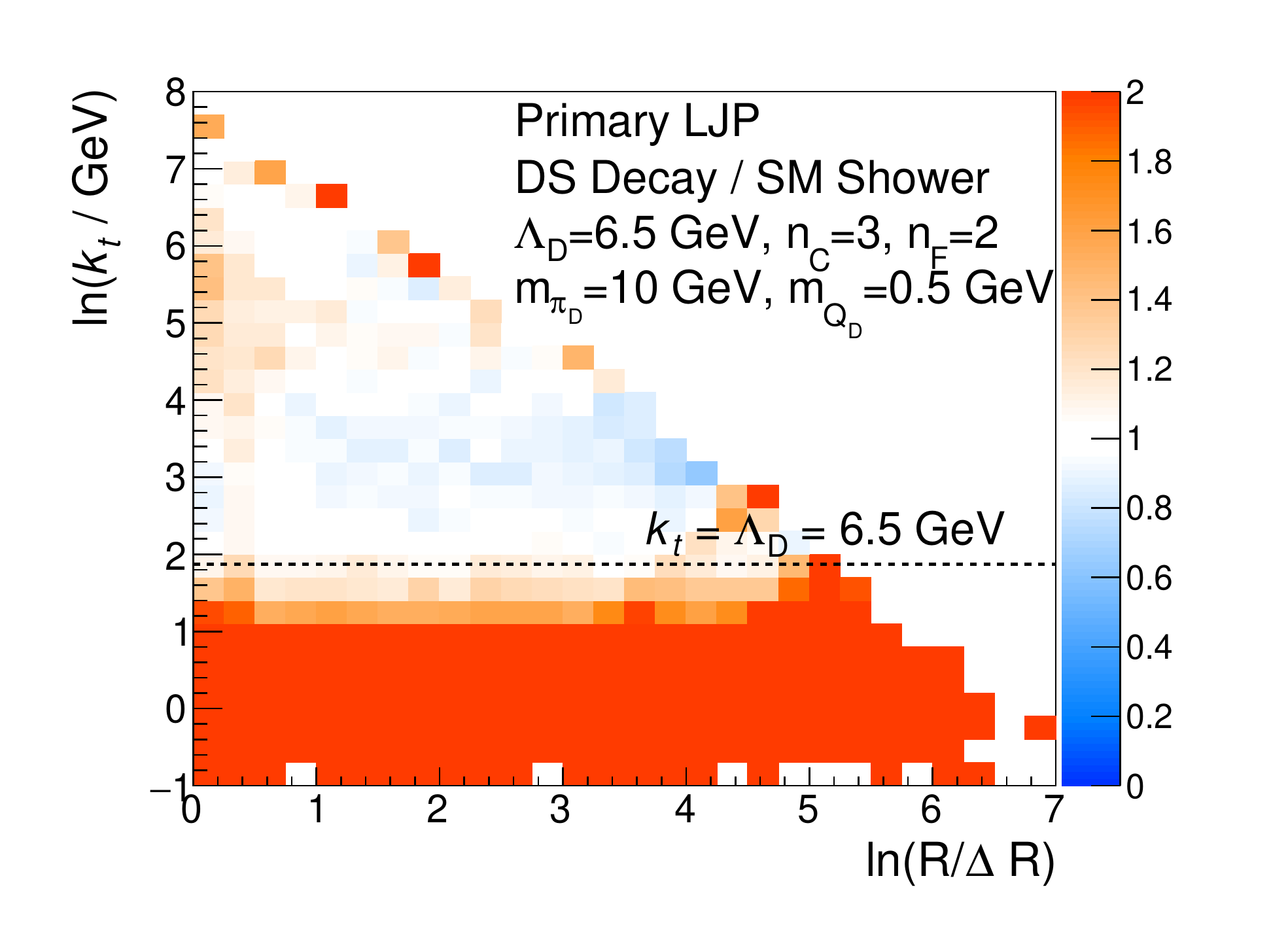} \\
  \includegraphics[width=0.38\textwidth]{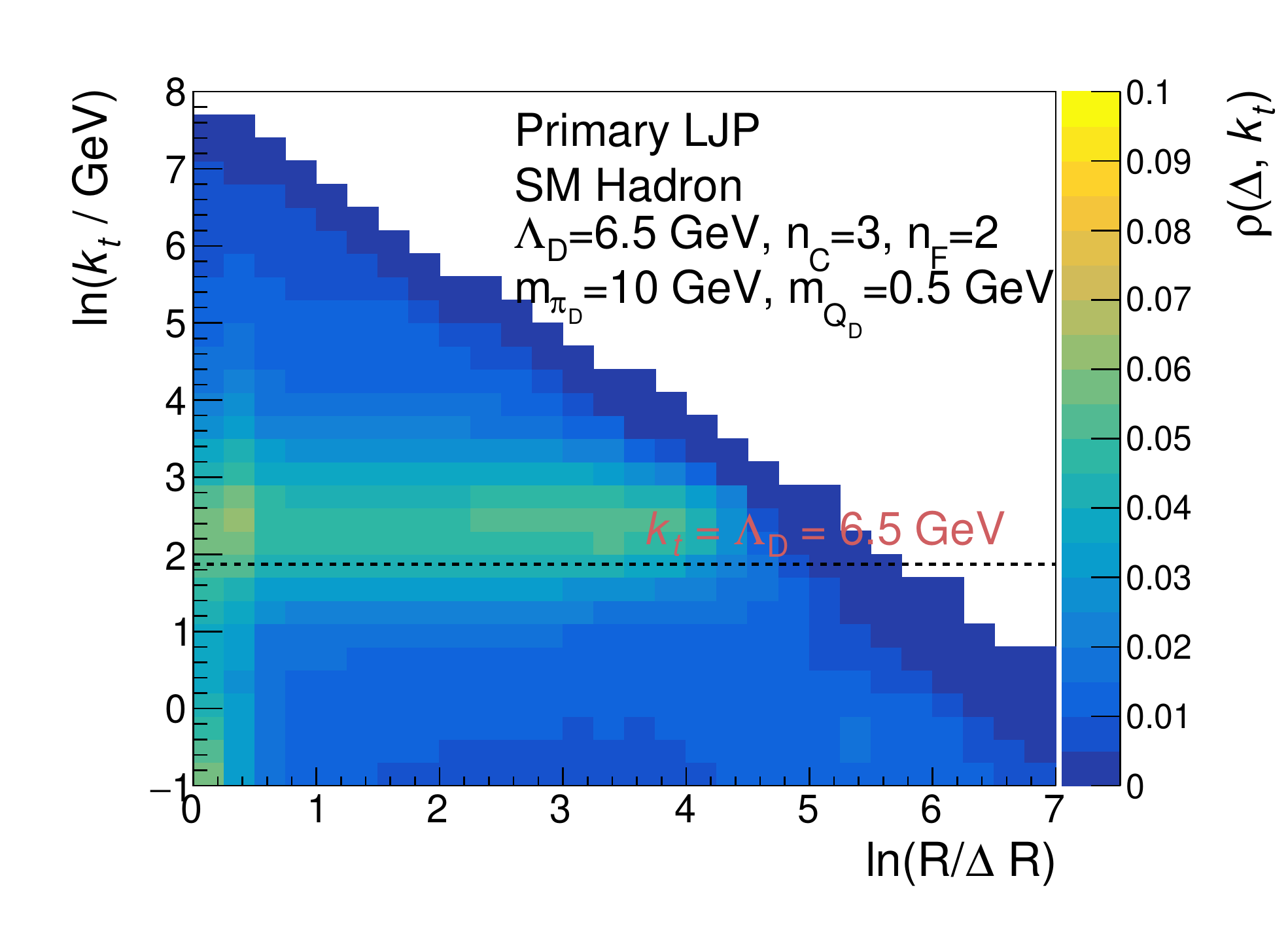} 
  \includegraphics[width=0.38\textwidth]{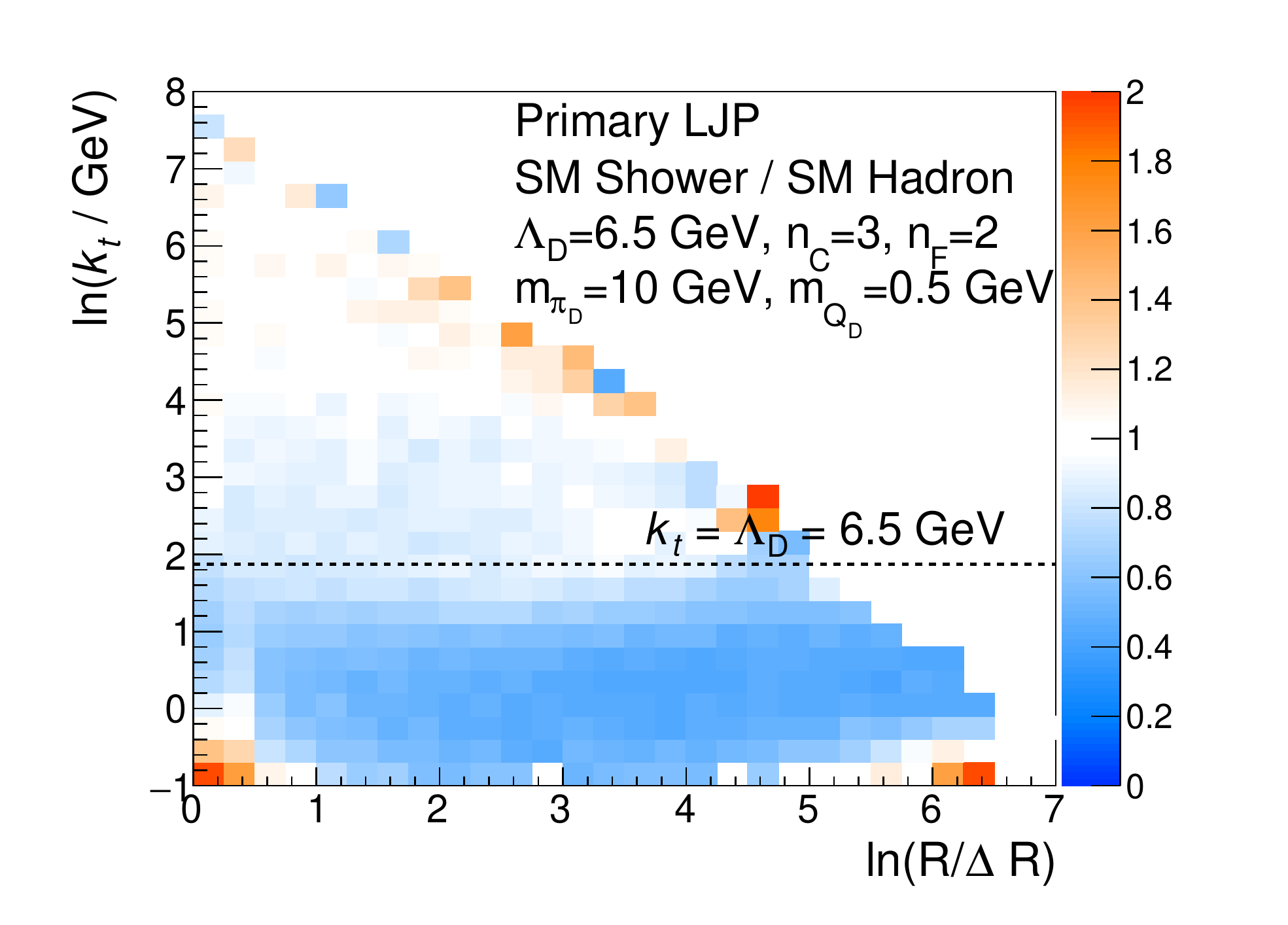} \\
  \caption{The left column provides the LJP for the following stages: (top) dark sector shower, (second) dark sector hadronization, (third) dark sector decay, (fourth) Standard Model shower, and (bottom) Standard Model hadronization.  The right column provides the ratio of the LJP for the following:(top) the dark shower to the dark hadron, (second) dark hadron to dark hadron decay, (third) dark hadron decay to Standard Model shower, and (bottom) the Standard Model shower to Standard Model hadron.}
  \label{fig:snowmass:LJP}
\end{figure}

\noindent
The Snowmass whitepaper on dark showers~\cite{Albouy:2022cin} proposed a variety of benchmark models.
These models incorporate input from lattice simulation to motivate a set of parameterized models of the dark hadron spectra.  They additionally incorporate decays among the dark hadrons.
Since the parameters of this model are slightly different than the baseline chosen for our simple benchmark, it is useful to check that the same conclusions apply to this model. The LJP for each stage is shown in \cref{fig:snowmass:LJP}. As expected, the same features are seen as compared to the benchmark provided in the main text.

\clearpage

\section{Impact of $\mathbf{\lambdaD}$}
\begin{figure}[thb]
  \centering
  \includegraphics[width=0.38\textwidth]{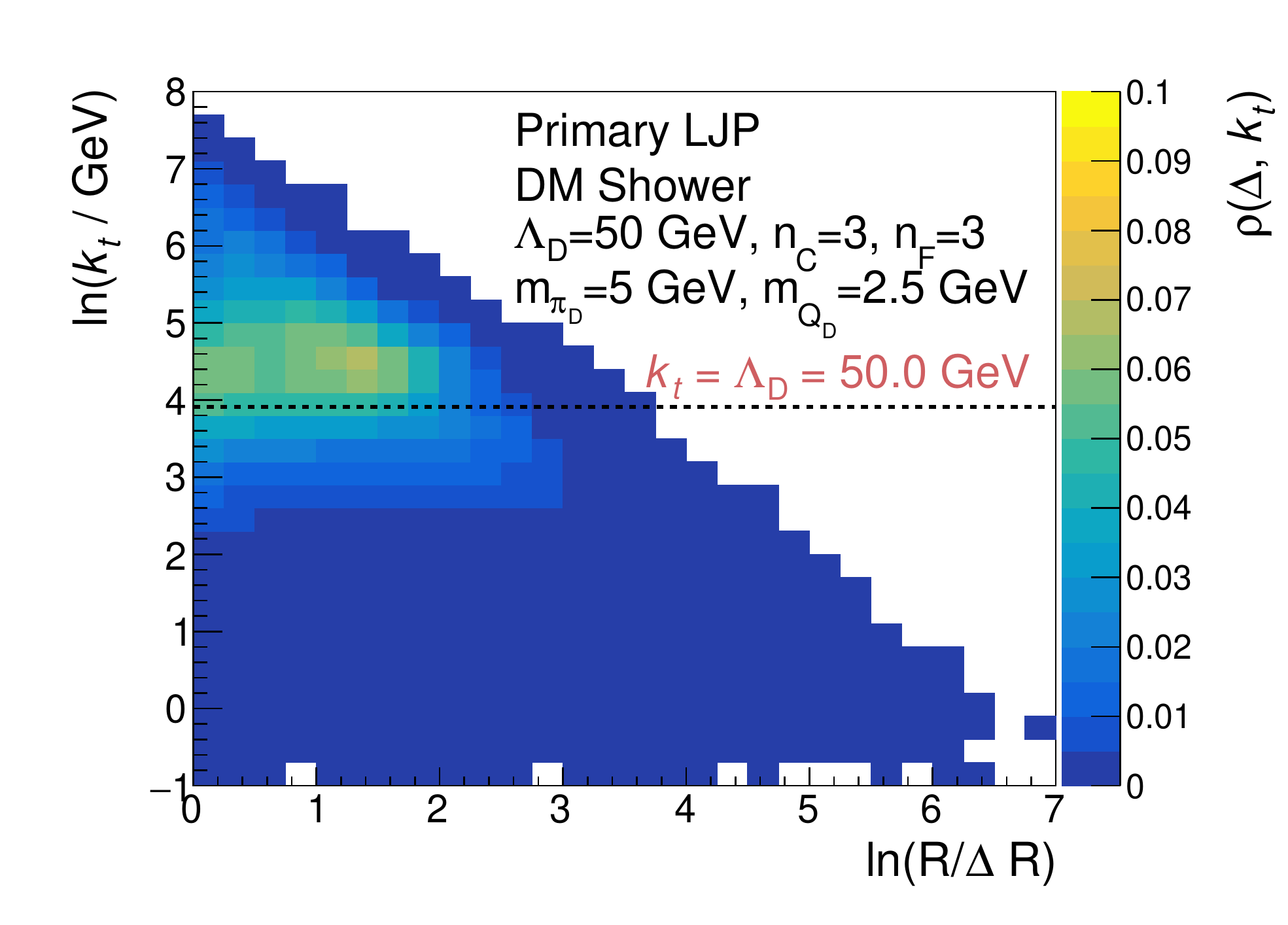}
  \leavevmode\phantom{\includegraphics[width=0.4\textwidth]{plots/lambda50Plots/Kt_Lambda50DMShower.pdf}}
  \includegraphics[width=0.38\textwidth]{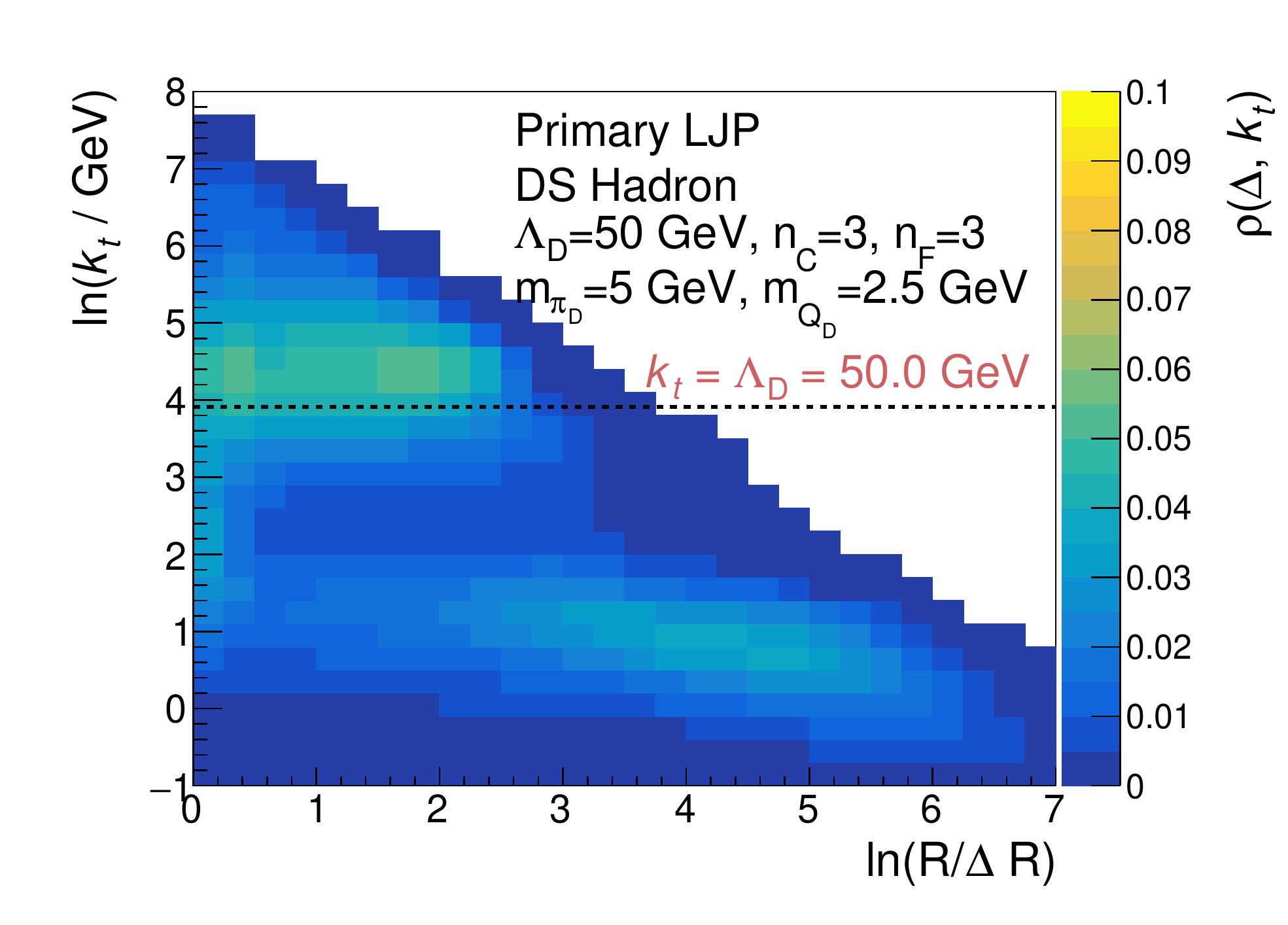}
  \includegraphics[width=0.38\textwidth]{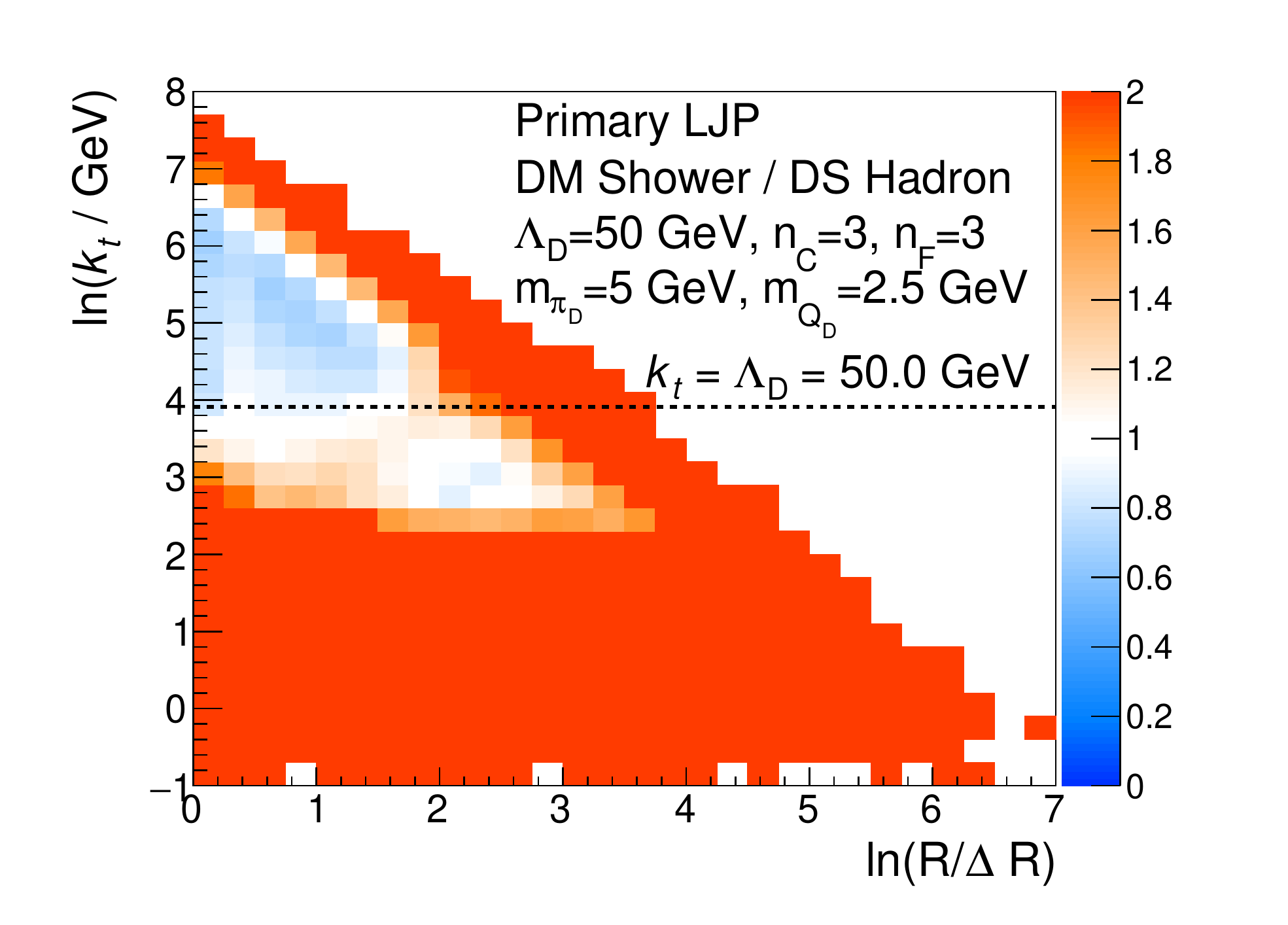} \\
  \includegraphics[width=0.38\textwidth]{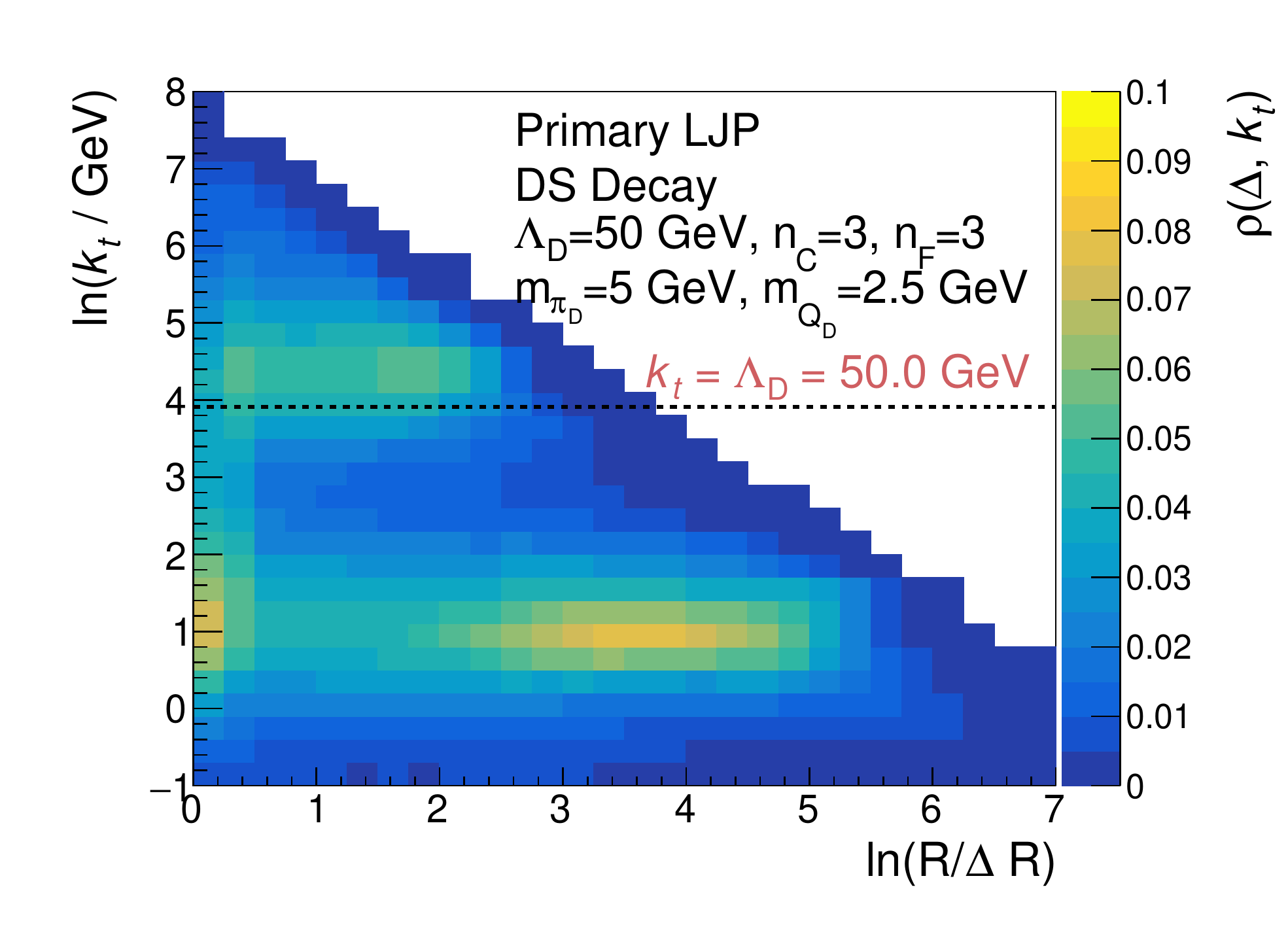}
  \includegraphics[width=0.38\textwidth]{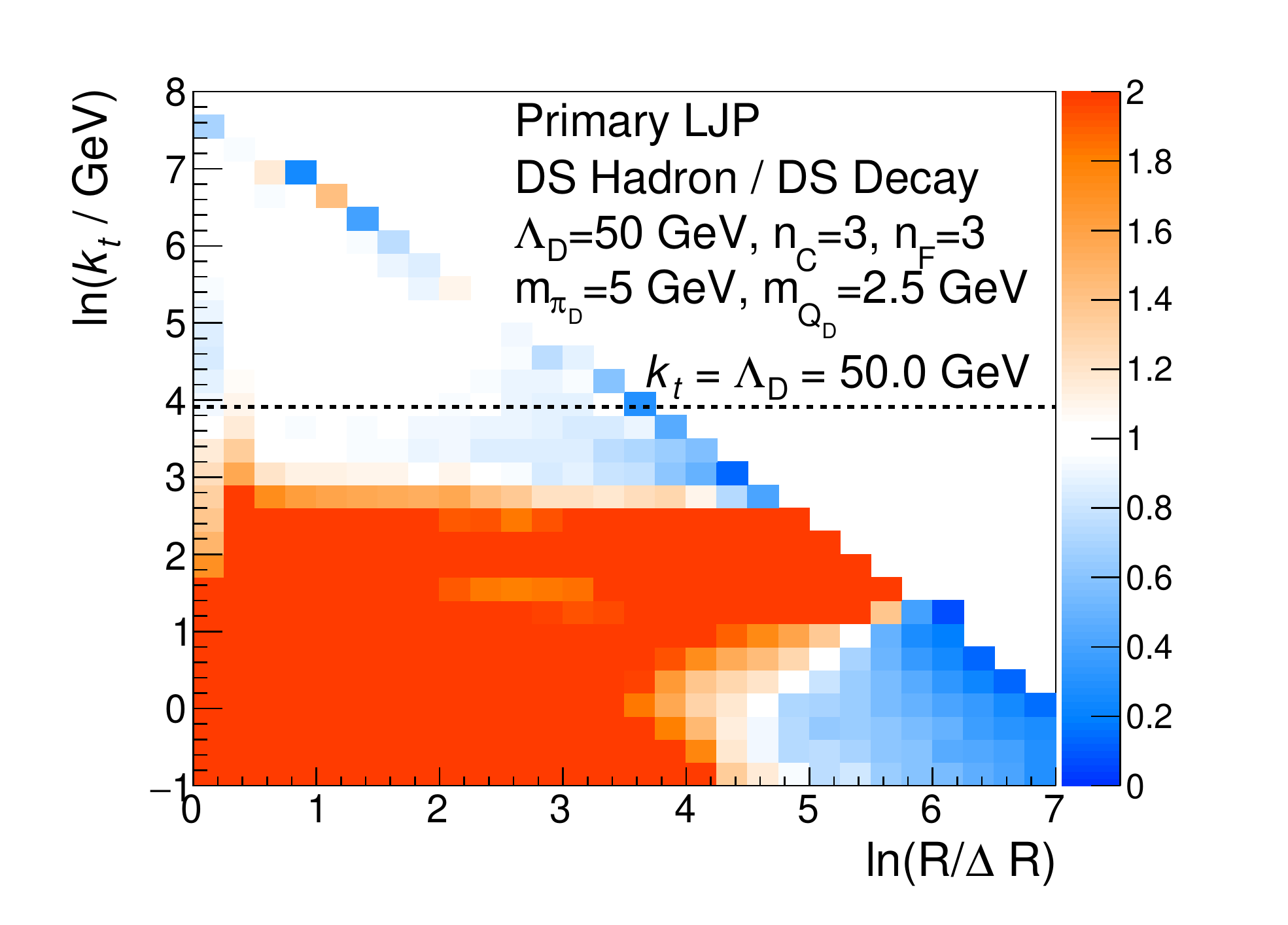} \\
  \includegraphics[width=0.38\textwidth]{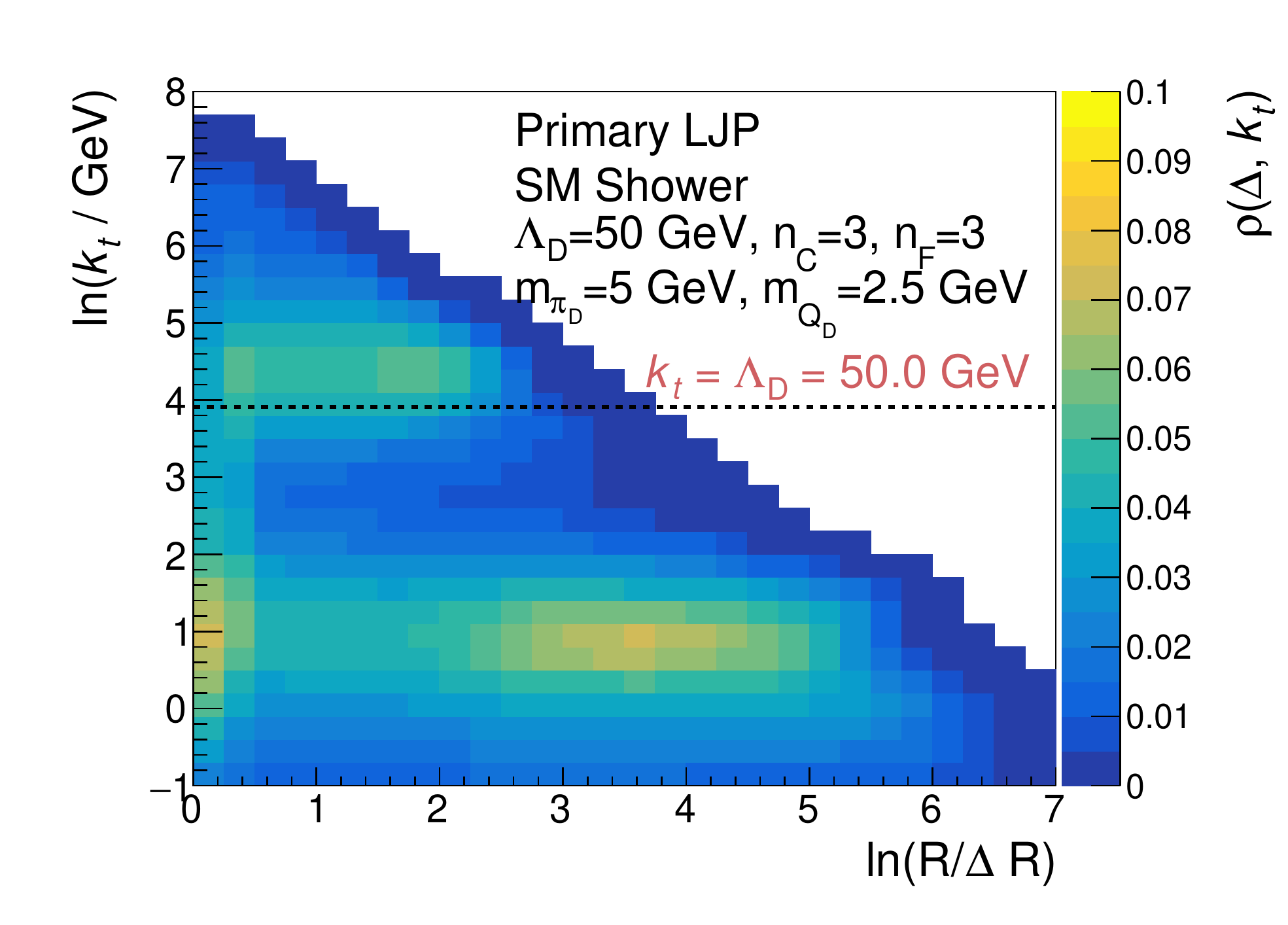}
  \includegraphics[width=0.38\textwidth]{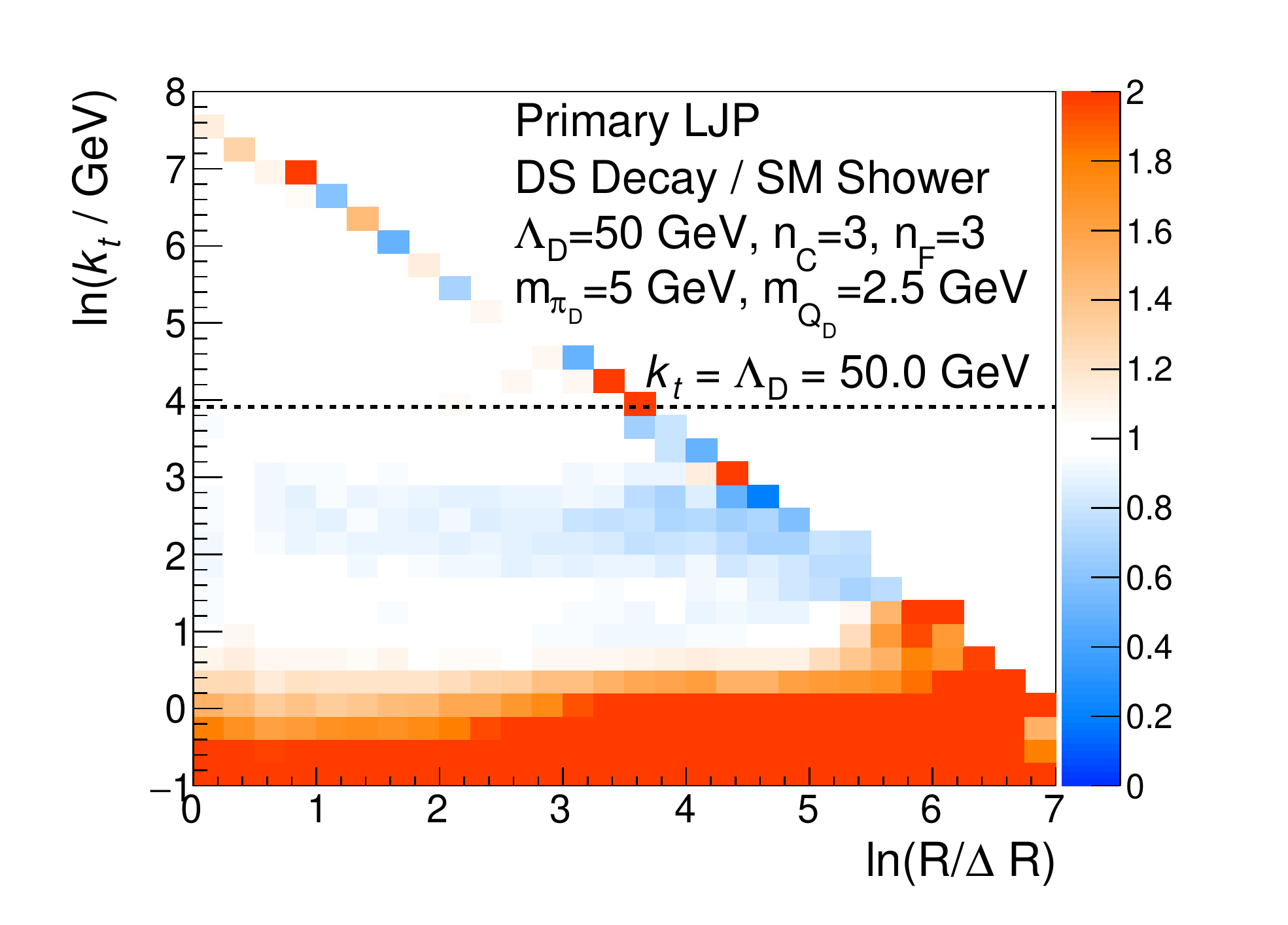} \\
  \includegraphics[width=0.38\textwidth]{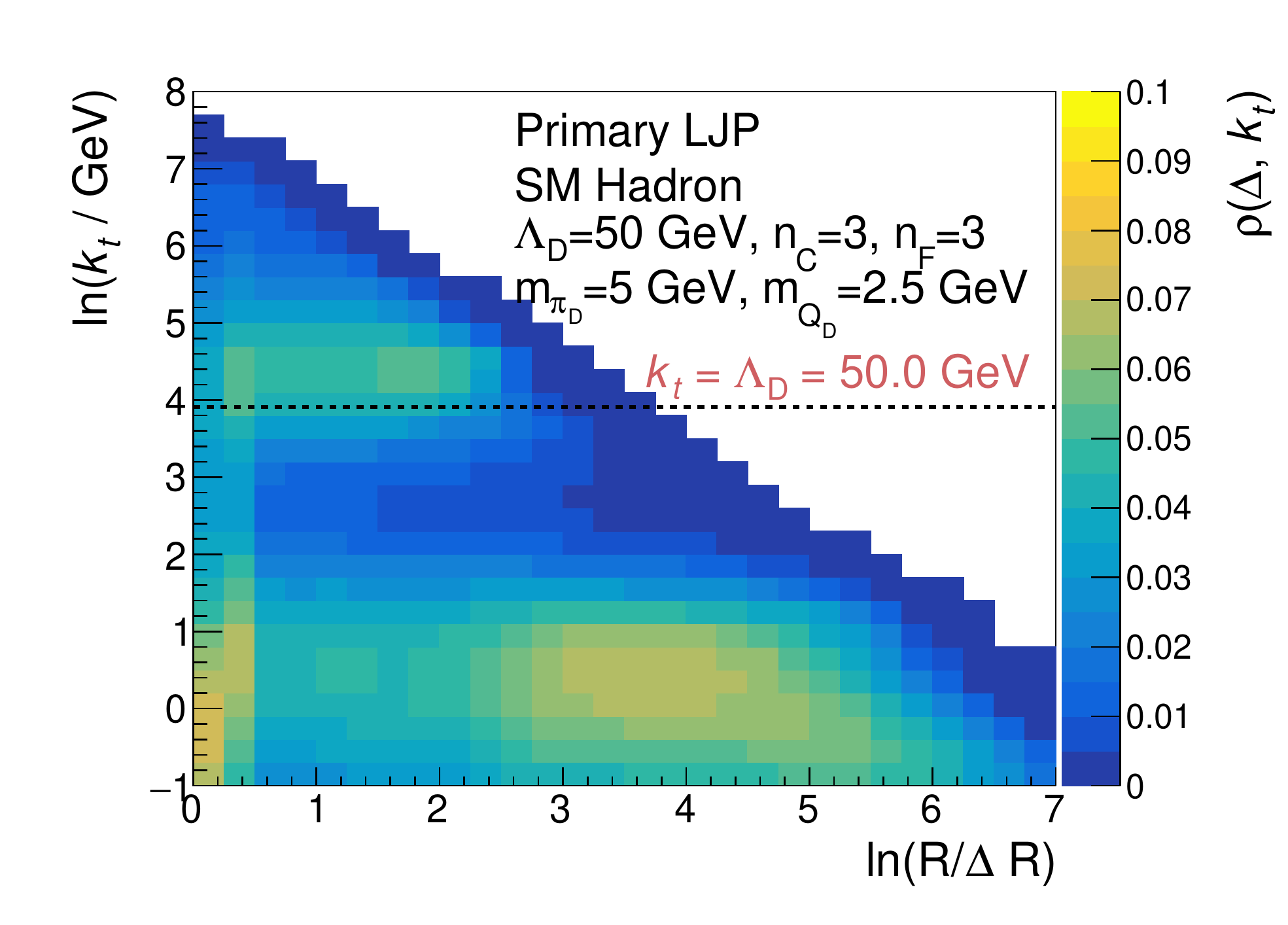} 
  \includegraphics[width=0.38\textwidth]{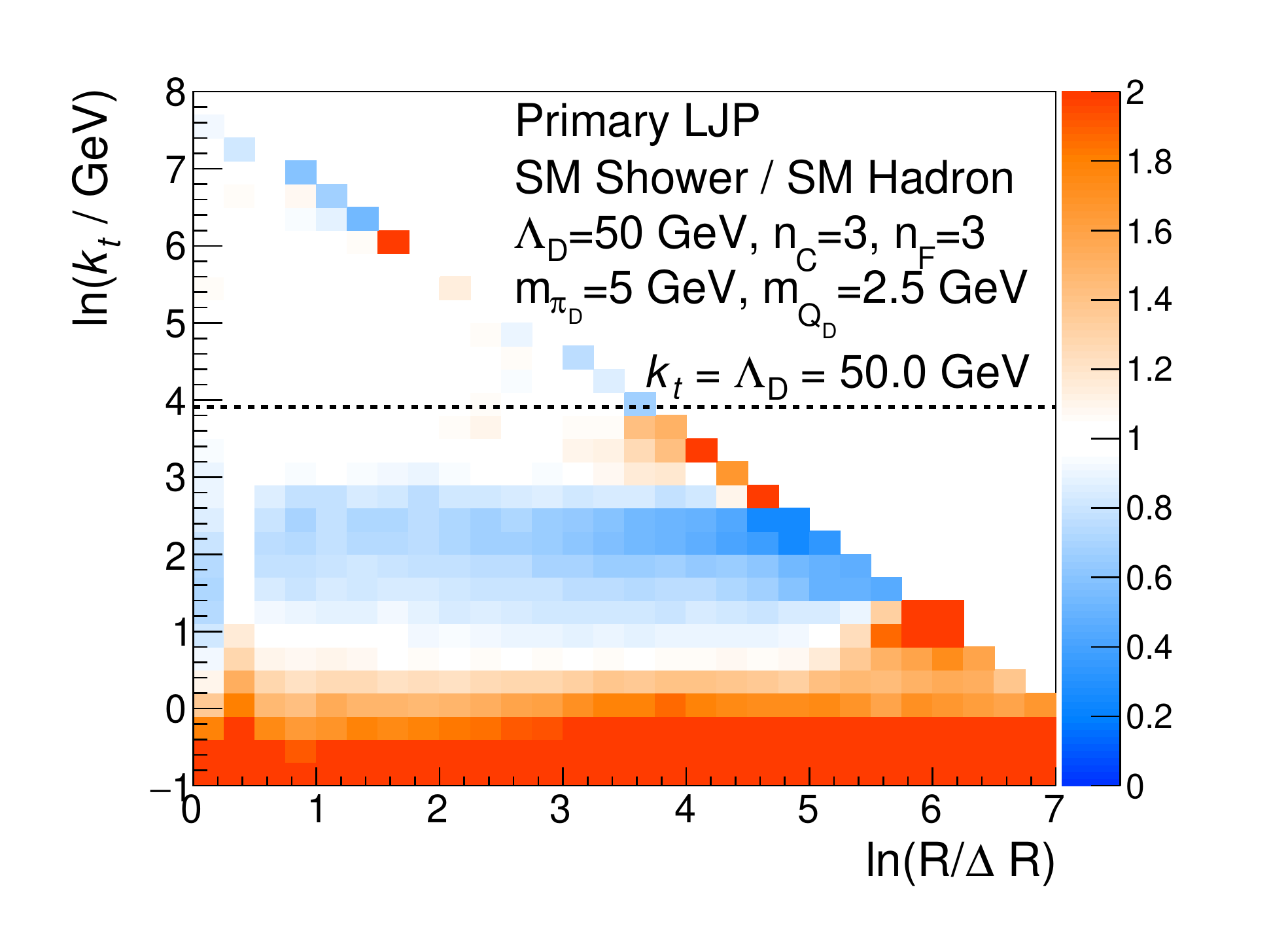} \\
  \caption{The left column provides the LJP for the following stages: (top) dark sector parton, (second) dark sector hadronization, (third) dark sector decay, (fourth) Standard Model shower, and (bottom) Standard Model hadronization.  The right column provides the ratio of the LJP for the following:(top) the dark shower to the dark hadron, (second) dark hadron to dark hadron decay, (third) dark hadron decay to Standard Model shower, and (bottom) the Standard Model shower to Standard Model hadron.}
  \label{fig:lambda50:LJP}
\end{figure}

\noindent
There are several free parameters in the dark showers model. Each of which results in various changes to the LJP. One particularly interesting parameter to consider is $\lambdaD$, which impacts the scale where the dark shower stops and hadronization occurs. The LJP after the dark shower is shown for $\lambdaD = 50~\text {GeV}$ in \cref{fig:lambda50:LJP}, which illustrates that the shower stops at much higher values of $k_{t}$, as expected.  This higher value of $\lambdaD$ results in new features in the LJP that were masked for lower values of $\lambdaD$. With the large gap between the meson mass and $\lambdaD$, a second band appears in the DS hadronization stage.  These features deserve further exploration.  It would be interesting to develop analytic tools to calculate the LJP for models with this wide separation of scales.  From a phenomenological perspective, it would be very interesting to design searches that take advantage of these features (or are explicitly agnostic to them), perhaps even using machine learning.

\end{document}